# Phonon transport and thermal conductivity in two-dimensional materials

Xiaokun Gu and Ronggui Yang

Department of Mechanical Engineering, University of Colorado, Boulder, CO, USA, 80309

ronggui.yang@colorado.edu

**Abstract**

Two-dimensional materials, such as graphene, boron nitride and transition metal dichalcogenides, have attracted increased interest due to their potential applications in electronics and optoelectronics. Thermal transport in two-dimensional materials could be quite different from three-dimensional bulk materials. This article reviews the progress on experimental measurements and theoretical modeling of phonon transport and thermal conductivity in two-dimensional materials. We focus our review on a few typical two-dimensional materials, including graphene, boron nitride, silicene, transition metal dichalcogenides, and black phosphorus. The effects of different physical factors, such as sample size, strain and defects, on thermal transport in Two-dimensional materials are summarized. We also discuss the environmental effect on the thermal transport of two-dimensional materials, such as substrate and when two-dimensional materials are presented in heterostructures and intercalated with inorganic components or organic molecules.

Nomenclature:

Constants:

$k_\text{B}$ : Boltzman constant, $1.38 \times 10^{-23}$ m²kg/s²

$\hbar$ : reduced Planck constant, $1.055 \times 10^{-34}$ Js

Symbols:

$a$ = ratio of temperature rise
$A$ = cross-sectional area, m²
$E$ = interatomic potential energy, J
$g$ = interfacial thermal conductance, W/K
$G$ = thermal conductance, W/K
$h$ = total energy, J
$J$, $\mathbf{J}$ = heat flux, W/m²
$L$ = dimension of sample along the direction of heat flux
$m$ = atomic mass, kg
$n^0$ = equilibrium phonon distribution function
$n$ = non-equilibrium phonon distribution function

$N$ = number of monolayers in a few-layer two-dimensional material
$P$ = heating power, W
$\delta P$ = absorbed heating power, W
$\dot{q}_h$ = volumetric heat generation power, W/m³
$\mathbf{q}$ = phonon wavevector
$r$ = distance along radius direction, m
$r_0$ = radius of a laser beam for thermal measurement, m
$r_a$ = radius of atom, m
$R$ = radius of a hole, where 2D sample is placed in Raman measurement, m
$\mathbf{R}$ = position vector, m
$s$ = phonon polarization index
$S$ = heat current, Wm
$t$ = thickness of two-dimensional material, m
$T$ = temperature, K
$\Delta T$ = temperature difference, K
$u$ = atom displacement, m
$v, \mathbf{v}$ = group velocity
$V$ = volume, m³
$W$ = width of nanoribbon, m
$x, y, z$ = coordinate

Greek symbols:
$\alpha$ = optical absorption coefficient
$\kappa$ = thermal conductivity, W/mK
$\tau$ = relaxation time, s
$\Gamma$ = scattering rate, s⁻¹
$\gamma$ = Gruneisen parameter
$\varepsilon$ = a constant parameter
$\phi$ = 2nd-order (harmonic) force constant, J/m²
$\psi$ = 3rd-order anharmonic force constant, J/m³
$\chi$ = temperature coefficient for Raman shift, K/cm⁻¹
$\omega$ = angular frequency, rad Hz
$\Delta\omega$ = frequency shift, rad Hz

Subscripts:
a = ambient
beam = beam
c = contact

calculated = calculated
heat = heating membrane
$i$ = atom index
in = intrinsic
m = location of electron beam
measured = measured
**q** = phonon wavevector
**R** = position vector
$s$ = phonon polarization index
sen = sensing membrane
sup = supported sample
sample = sample
$\tau$ = index for basis atom

Superscripts:
$i$ = index of scattering
B = boundary
D = defect
V = vacancy
$x, y, z$ = coordinate
$\alpha, \beta, \gamma$ = $x$, $y$ or $z$ direction
', ", ''' = atom index

Abbreviations:
2D: two-dimensional
AGF: atomistic Green's function
EMD: equilibrium molecular dynamics
h-BN: hexagonal boron nitride
LA: longitudinal acoustic
LO: longitudinal optical
PBTE: Peierls-Boltzman transport equation
MD: molecular dynamics
NEMD: non-equilibrium molecular dynamics
TA: transverse acoustic
TMD: transition metal dichalcogenide
TO: transverse optical
ZA: flexural acoustic
ZO: flexural optical

# 1 Introduction

Since its discovery in 2004 by Novoselov *et al*. [1, 2], graphene has attracted intensive attention due to its unique physical properties and the potential technological applications in electronics, photonics and many other fields [3, 4]. Inspired by the success of graphene, many other two-dimensional (2D) materials [5, 6], such as hexagonal boron nitride (h-BN), silicene, transition metal dichalcogenides (TMDs), transition metal oxides, five-layered V-VI trichalcogenides and black phosphorus, have been synthesized and studied. Figure 1 shows the lattice structures of some typical 2D materials. Many of these 2D materials have been shown to have similar or even superior properties than graphene. For example, while pristine graphene has no bandgap, some monolayer transition metal dichalcogenides, including $MoS_2$ and $WS_2$, exhibit direct bandgap, making them ideal for a wide range of applications in electronics and optoelectronics [7, 8]. Compared to conventional silicon-based electronics, these 2D material-enabled devices promise to be more efficient due to their smaller size. However, heat dissipation of these devices could become a bottleneck limiting their performance and reliability as thermal conductivity of many of these 2D materials could be very low. Understanding phonon transport and thermal conductivities in the 2D crystals could be very important for the design of novel devices using 2D materials.

Significant progress has been made on understanding nanoscale heat transfer over the past two decades with the focus on nanostructures, such as nanowires [9], superlattices [10] and nanocomposites [11, 12]. In these interface-dominant nanomaterials, interfaces play a crucial role on deviating thermal transport mechanisms from bulk materials, either through inducing an interfacial thermal resistance or the formation of new phonon bands [13]. The thermal conductivity of these nanostructures could be significantly lowered compared with their bulk counterparts, which leads to various applications, such as thermoelectrics [14], thermal insulation [15] and thermal protection [16]. On the other hand, the surface of the ultrathin 2D materials does not necessarily hinder the thermal transport by scattering energy carriers due to its atomic smoothness, which results in very different phonon transport and thermal conductivity in 2D materials from conventional thin films.

There have been quite some studies on thermal transport in 2D materials. As the first 2D material that gained overwhelming attention, graphene has been intensively studied in the past few years [17-20]. With a very high thermal conductivity in the range of 2000-5000 W/mK at room temperature, even higher than diamond, many works have been centered around understanding the thermal transport mechanisms and exploiting the applications of these superior heat-conducting materials. Apparently those studies on phonon transport and thermal conductivity of graphene have laid a great foundation for other emerging 2D materials. However, there are still quite some issues that the consensus has not been reached by researchers, even for graphene, the most well-studied 2D material. In addition, the crystal structures of many novel 2D materials are different from the "one-atom-thick" graphene, making the direct deduction of the knowledge from graphene to other 2D materials questionable.

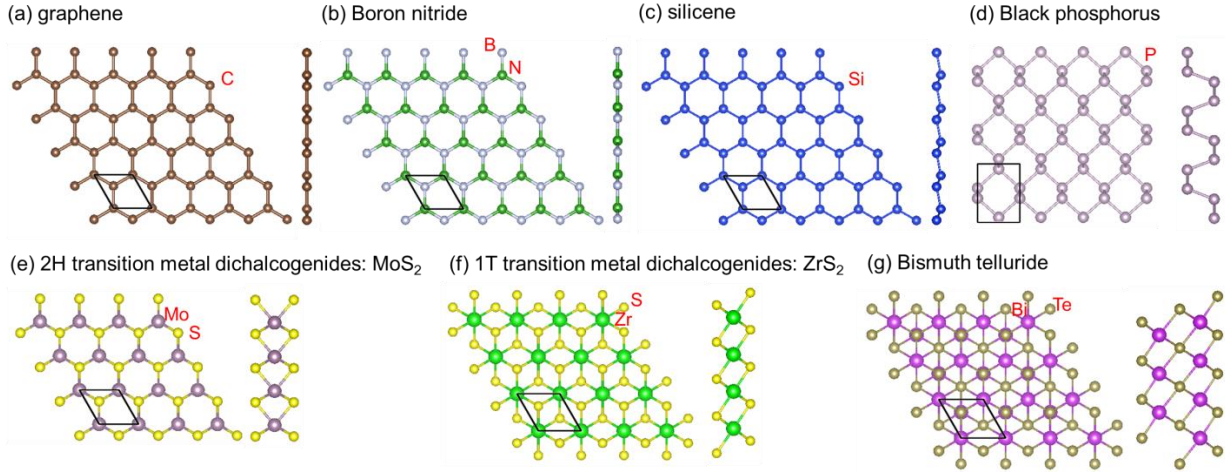

FIG. 1: Top view and side view of some typical 2D materials. The primitive unit cell of each material is indicated by black lines in the top view.

Many geometrical constraints and physical factors can be used to change the thermal conductivity of 2D materials, which could lead to new applications in thermal management and energy conversion. For example, the thickness and the feature size in the basal plane of 2D material thin films can be tuned, which in turn can be used to control thermal transport of 2D materials. Defects and mechanical strains can also be utilized to manipulate the thermal conductivity. Furthermore, thermal transport in 2D materials is highly dependent on their interaction with the environment. For example, the coupling with substrates and other 2D crystals in heterostructures could significantly alter phonon transport in 2D materials. Surface functionalization and intercalation also significantly alter thermal transport of 2D materials. Figure 2 summarize the geometric and physical factors that have been exploited to control the thermal conductivity of 2D materials which will be reviewed in this paper.

In this review, we focus on phonon transport and the lattice thermal conductivity of 2D materials In Section 2 and 3, we discuss the measurement techniques and atomistic simulation tools that are capable and often used for the measurement of the thermal conductivity and the modeling of phonon transport in 2D materials, respectively. Some of the thermal conductivity results for a few typical 2D materials are summarized in Section 4. In Section 5, we present the geometrical size effects on the thermal conductivity of 2D crystals while the effect of mechanical strains and defects on the thermal transport is presented in Section 6. Thermal transport of 2D materials in device geometry, such as the effect of substrates and heterostructured 2D materials, is summarized in Section 7. In Section 8, we discuss the strategies to alter the thermal properties of 2D materials through chemical tuning. We also note that there exist quite a few excellent reviews on the thermal properties of graphene [17-20]. In this review, we will touch on graphene just as a context to highlight the properties of other 2D materials.

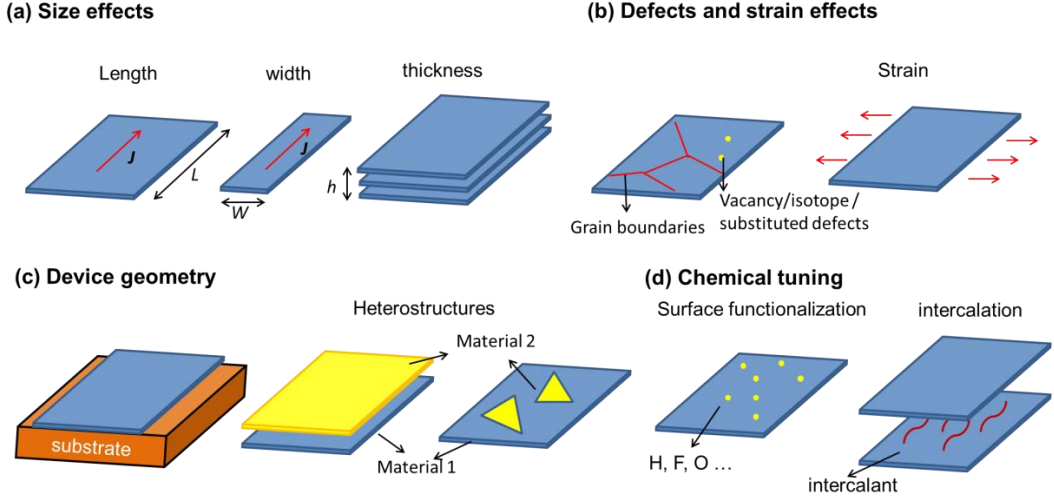

FIG 2: The geometrical and physical factors that affect the thermal conductivity of 2D materials.

## 2 Measurement techniques

Great progress has been made over the past two decades on the thermal conductivity measurement of thin films, especially the development of 3ω-method [21, 22] and the pump-probe thermoreflectance technique [23, 24]. In the 3ω-method, metal lines serving as both heating elements and resistive temperature sensor are deposited on the surface of the sample. Similarly, the samples for the pump-probe measurement are usually coated with metal thin film with a thickness of tens of nanometers serving as a heater and transducer whose optical properties changes with temperature. Such sample post-processing makes it extremely difficult, if not impossible, to measure the intrinsic thermal conductivity of 2D crystals, typically with a thickness in the order of 1 nm. On the other hand, the optothermal Raman technique [25] and the micro-bridge method [26] have been adapted to measure the thermal conductivity of 2D materials. As the measurement principles and the applications of these methods have been reviewed elsewhere, here we present a brief introduction of these techniques with an emphasis on the limitations and the possible origins of the measurement errors just for better understanding the measurement results presented in Section 4-8.

### 2.1 Optothermal Raman technique

A Raman spectroscopy-based thermal conductivity measurement on single-layer graphene was reported by Balandin *et al*. [25] in 2008. It was then applied to other 2D materials, such as multilayer graphene [27], isotopically modified graphene [28] and $MoS_2$ [29]. The schematics of the measurement system are illustrated in Fig. 3(a). This method is based on temperature dependence of the Raman spectra of 2D materials. Taking graphene as an example, the G peak position [25, 30], the 2D peak position [28, 31] or Stokes-to-anti-Stokes ratio [32] of Raman signal have been chosen as the signals for temperature detection. To quantify the relationship between Raman signals and temperature, a series of Raman spectroscopy measurements were carried out first by changing the sample temperature before thermal conductivity measurement. As an example, Fig. 1(b) shows the peak positions of the G peak and the 2D peak of Raman signal in a single-layer graphene at different temperature, which are found to have linear relationship. The ratio between the temperature change and the peak frequency shift is denoted as a temperature-independent coefficient $\chi = \Delta T / \Delta \omega$.

When conducting the thermal conductivity measurement, a 2D material sample is suspended above holes or trenches microfabricated, and then heated by a focused laser light with the heating power $P$, which induces the temperature rise of the sample. The temperature profile is illustrated in Fig. 3(a). The averaged local temperature rise, $\Delta T_{measured}$, within the laser spot can be read from the change of Raman signal using the pre-measured known relationship between Raman signal and temperature. For example, $\Delta T = \chi \Delta \omega$ for Raman G peak or Raman 2D peak in graphene, as shown in Fig. 3(b) [33]. With the temperature increase derived, the thermal conductivity of the sample can then be calculated from heat conduction model as follows [30].

When the sample is suspended above a hole with a radius of $R$ and an effective thermal conductivity $\kappa$ is assumed for the entire suspended sample, the temperature $T(r)$ at a point whose radial position from the center of the hole is $r$, is governed by the heat conduction equation as

$$\kappa \frac{1}{r}\frac{d}{dr}\left[r\frac{dT(r)}{dr}\right] + \dot{q}_h(r) = 0 \text{ for } r \leq R, \tag{1a}$$

$$\kappa_{sup} \frac{1}{r}\frac{d}{dr}\left[r\frac{dT(r)}{dr}\right] - \frac{g}{t}[T(r) - T_a] = 0 \text{ for } r \geq R \tag{1b}$$

with the boundary condition:

$$T(\infty) = T_a. \tag{1c}$$

where $T_a$ is the temperature of the substrate which is assumed to be in equilibrium with the measurement chamber. Because the thermal conductivity of the sample in contact with the support could be much lower than the suspended one due to phonon leakage effect [34] when the 2D crystal is placed on the substrate, for example, the thermal conductivity of the supported graphene is only about 1/5 of that of the suspended graphene [34], we use single $\kappa_{sup}$ to represent the thermal conductivity of the portion of the 2D material being supported and write the heat conduction model in Eq.(1) into two parts. Aside from the substrate-induced reduction on thermal conductivity of 2D material, there could be a temperature jump between the sample and the substrate, and we use $g$ to denote the interfacial thermal conductance between the sample and the substrate.

Assuming $t$ is the thickness of the 2D material, $\delta P$ is the fraction of the absorbed heating power, the volumetric optical heating with the radius of the laser beam $r_0$ can be written as

$\dot{q}_h(r) = \frac{\delta P}{\pi r_0^2 t} \exp\left(-\frac{r^2}{r_0^2}\right)$ assuming that the transmitted laser light is not reflected back to the sample by the substrate beneath the suspended sample, $\delta P$ equals $\alpha \cdot P$, with the optical absorption coefficient $\alpha$.

If $\kappa_{sup}$ and $g$ are known (by other measurements), the temperature profile $T(r)$ could be determined by a single valuable $\alpha \cdot P / \pi r_o^2 h \kappa$. After solving Eq. (1) for $T(r)$, the averaged local temperature at the laser spot can be approximated by

$$T_{calculated} = \frac{\int_0^\infty T(r)\dot{q}_h(r)rdr}{\int_0^\infty \dot{q}_h(r)rdr}. \tag{2}$$

The thermal conductivity of the 2D material sample can then be determined by adjusting the value of thermal conductivity to match $T_{calculated} - T_a = \Delta T_{measured}$. For more complicated geometries, other than circular holes, numerical methods, such as finite element analysis, can be used to solve for the temperature profile.

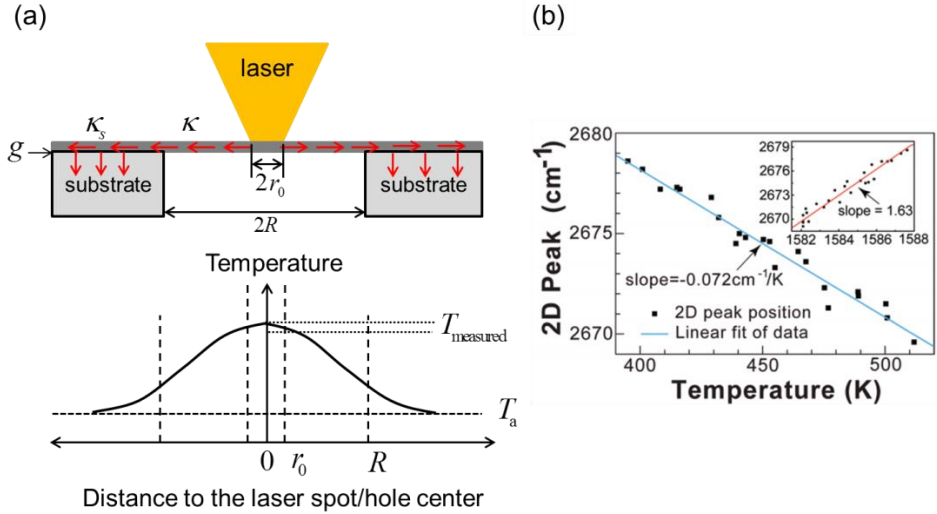

FIG. 3: (a) Schematic of sample geometry for the thermal conductivity measuremet using optothermal Raman technique. (b) The measured relationship between temperature and the shift of the 2D peak and G peak (in the inset) of the Raman signal of a single-layer graphene. From Ref [33] (copyright 2010, American chemical society).

There are several mechanisms that could contribute to experimental errors. Since only a small portion of laser power is absorbed, which is then conducted by the 2D material, the accuracy of the measured thermal conductivity is directly connected with the value of optical absorption used, as indicated in Eq. (1a). The values of optical absorption employed to derive the thermal conductivity of graphene ranged from 2.3% to ~11% [25, 30, 32]. The scattered value of the reported thermal conductivity of graphene could be partially attributed to the value used for optical absorption coefficient.

Another possible source of the experiment errors comes from the values for the interfacial thermal conductance between the 2D material sample and the substrate, $g$, and the thermal conductivity of the supported 2D samples, $\kappa_{sup}$. Since heat flows across the suspended sample and is eventually dissipated by contacting with the substrate, the accurate determination of interfacial thermal conductance and the thermal conductivity of the supported 2D materials plays an important role. Some works have adopted the values of $\kappa_s$ and $g$ from other independent experiments while others have even neglected the effects of $g$ and $\kappa_s$. Two methods have been proposed to avoid the measurement error from the substrate effects. Cai *et al*. [30] proposed to directly measure both $g$ and $\kappa_{sup}$ from an independent optothermal Raman experiment, where the 2D material completely supported by substrate is heated by the laser. Since the temperature increase of the heating region is only determined by $g$ and $\kappa_{sup}$, it is possible to extract these two quantities using the similar approach as is done for the suspended sample. Reparaz *et al*. [35] performed temperature mapping using two laser beams: one is used to create a thermal field, and the other probes the local temperature through the spectral position of a Raman active mode. The thermal conductivity of the suspended sample can then be obtained by fitting the slope of the spatial decay of thermal field in the suspended sample without of the input need of $\kappa_{sup}$ and $g$.

The size of the laser beam and the size of the holes or trench suspending the sample might also lead to discrepancies. As the spot size used in Raman measurements is in the order of hundreds of nanometers, which could be much shorter than the mean free path of some phonons in 2D materials [36], high thermal conductivity value of some of the 2D materials might be underestimated. In addition, if the distance between the heat source (laser spot) and the heat sink (substrate) is small, some phonons can ballistically transport across the sample, making the thermal conductivity sensitive to the sample size.

In addition to the experiment errors discussed above, there are a few sources of measurement uncertainty that need to be taken into account when interpreting or reporting thermal conductivity data, including the uncertainty of the optical absorption coefficient $\alpha$, the uncertainty of the coefficient $\chi$ when quantifying the relationship between Raman signals and temperature, and the uncertainty of the peak position shift $\Delta\omega$ read from Raman Raman spectroscopy in the temperature measurement. Combining these sources of uncertainty, the uncertainty of the measured thermal conductivity could be as large as 40% in optothermal Raman experiments [18]. Even though the experiment uncertainty is large in optothermal Raman measurement, it is still one of the most simple and efficient way to measure the thermal conductivity of 2D materials, which is increasing used.

## 2.2  Micro-bridge method

The micro-bridge method [37] was originally developed by Shi *et al*. to measure the thermal conductivity of one-dimensional (1D) nanostructures, such as carbon nanotubes [38] and nanowires [9]. The technique has been extended to the thermal conductivity measurement of both suspended [26] and supported 2D materials [34]. Figure 4 illustrates a typical experiment setup.

The measurement sample such as carbon nanotubes, nanowire, or 2D materials, is placed on top of the two microfabricated silicon nitride or silicon dioxide membrane islands which are suspended several microns apart, each of which has metal resistor underneath. A direct current is applied to one of the two metal resistors which raises the temperature of the suspended island due to Joule heating. The heat current, $J$, flows across the sample and leads to a temperature increase in the other island and eventually dissipates to the environment through beams used to support the sensing membrane. The metal resistor at each island is also connected with four Pt leads, allowing four-probe electrical resistance measurement. The temperature can be read from the electrical resistance change. With the temperature measured, $T_{heat}$ for heating membrane and $T_{sen}$ for sensing membrane, the thermal conductance across the sample, $G_{sample}$, can be extracted through thermal resistance circuit analysis as,

$$J = G_{beam}(T_{sen} - T_a) = G_{sample}(T_{heat} - T_{sen}) \tag{3}$$

where $G_{beam}$ is the total thermal conductance of the five beams connecting to the sensing membrane, which can be determined by the thermal conductivity and dimensions of the beams. The thermal resistance across the sample, $1/G_{sample}$, comes from two resources: one is the contact resistance, $1/G_c$, and the other the intrinsic thermal conductance of the sample, $1/G_{in} = L/A\kappa$, with the cross-section area $A$ and sample length $L$, through

$$1/G_{sample} = 1/G_c + L/A\kappa \tag{4}$$

It is generally difficult to extract the contributions from contact resistance and internal (intrinsic) thermal resistance of the measured sample from the total conductance from a single measurement. One solution is to neglect the contact resistance if its contribution is expected to be small. For example, the contact resistance is estimated to less than 5-6% of the total measured resistance across a bilayer graphene with a length of 5 μm [26]. In some cases, the contact resistance is more important. In order to extract both the contact resistance and the thermal conductivity of the sample through Eq. (4), measurements with two or more sample with different sizes are usually required [39].

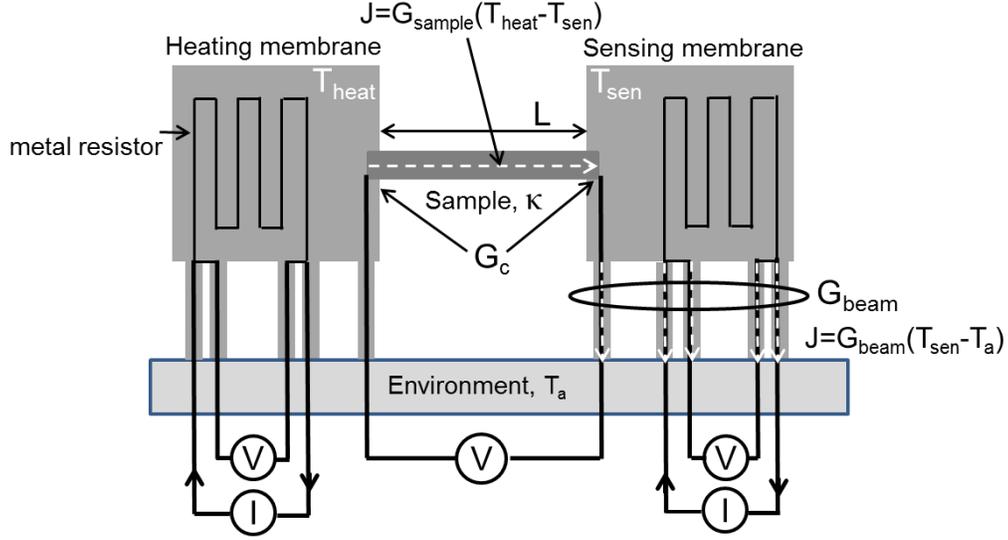

FIG. 4. Schematic of the thermal conductivity measurement through micro-bridge technique with a heater-sensor setup. The electric current is applied to the resistor in the heating membrane leading to temperature rise in the heating membrane. The heat current, indicated by the write arrows, flows across the sample and is dissipated to environment through the five beams. The voltage meter and current meter at each side are used to measure the electrical resistance of the metal resistor, through which the temperature can be read. The voltage meter in the middle is used to measure the electric voltage due to thermoelectric effect, which is irrelevant to the thermal conductivity measurement discussed in this paper. Adopted from [37].

To simultaneously measure both the intrinsic thermal conductivity of the sample and contact resistance from a single sample measurement, recently another method using a focused electron beam in a scanning electron microscopy (SEM) to induce localized heating in the sample has been proposed [40, 41]. Figure 5 shows the measurement principle and the setup which is compatible with the conventional micro-bridge method. By moving the electron beam can be moved, the heating location is changed leading to different temperature change at the two ends. By measuring the temperature rise at heater membrane $\Delta T_{heat}(=T_{heat}-T_a)$ and at sensor membrane $\Delta T_{sen}(=T_{sen}-T_a)$, the thermal conductance from the heating position to the heater membrane $G_m$ is $G_{beam}(1+a_m)/(a_0-a_m)$. Here $a_m$ is the ratio of temperature rise $\Delta T_{heat,m}/\Delta T_{sen,m}$ when the electron beam is at location $m$, and $a_0$ is the ratio of temperature rise when the electron beam is at the heater membrane. Both the contact resistance and thermal conductivity can then be extracted from a single measurement from the relation between $G_m$ and $m$.

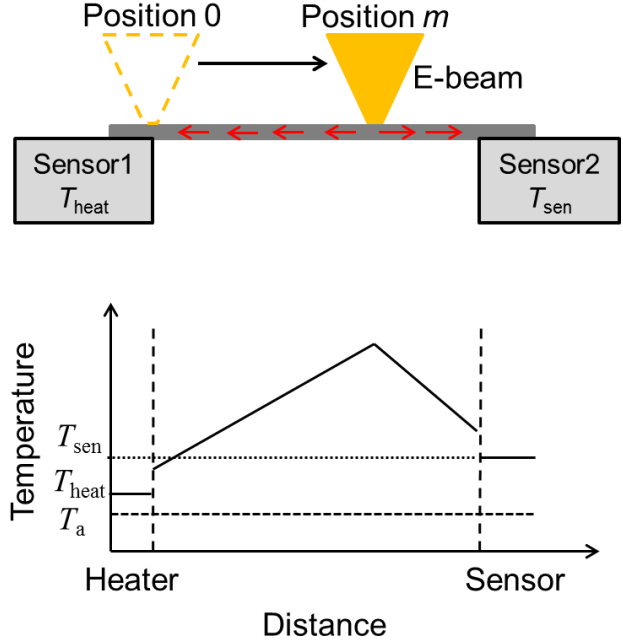

FIG. 5: Schematic of the thermal conductivity measurement through micro-bridge technique using a moving electron beam (E-beam) to heat the sample and two sensors to measure the temperatures at the two ends of the sample. Heat current is indicated by red arrows.

## 3 Theoretical modeling

Phonons, the energy quanta of lattice vibrations, are the dominant heat carriers in semiconductors and insulators. We focus our review on the modeling techniques for phonon transport in nonmetallic 2D materials. Due to the simple crystal structure of 2D materials and the increased computational power, many atomistic simulations find their great position for understanding phonon transport and thermal conductivity of 2D materials. To study thermal transport in a crystal with periodic lattice structure, one can model the lattice vibration and thermal conductivity from either real space or reciprocal space. In the real space approach, such as molecular dynamics (MD) simulations [42, 43], the thermal conductivity of the crystal is obtained by monitoring the movement of each atom. In the reciprocal space approach, the lattice vibration is decomposed to normal modes *i.e.*, phonons in the quantum mechanics point of view. The contribution of each phonon mode to the thermal conductivity is then determined by the calculation of phonon dispersion and phonon scattering matrix. The thermal conductivity can then be calculated using the kinetic theory such as Peierls-Boltzmann transport equation (PBTE) based method [44, 45]. When elastic scattering is dominant (at low temperature), atomistic Green's function (AGF) approach has been regarded as an efficient tool to model the ballistic or semi-ballistic phonon transport [46-48]. In either approaches, the potentials or force-fields that describe the interatomic interactions are essential. Figure 6 summarizes the atomistic simulation methods that have been used to calculate phonon properties and thermal conductivity of 2D materials.

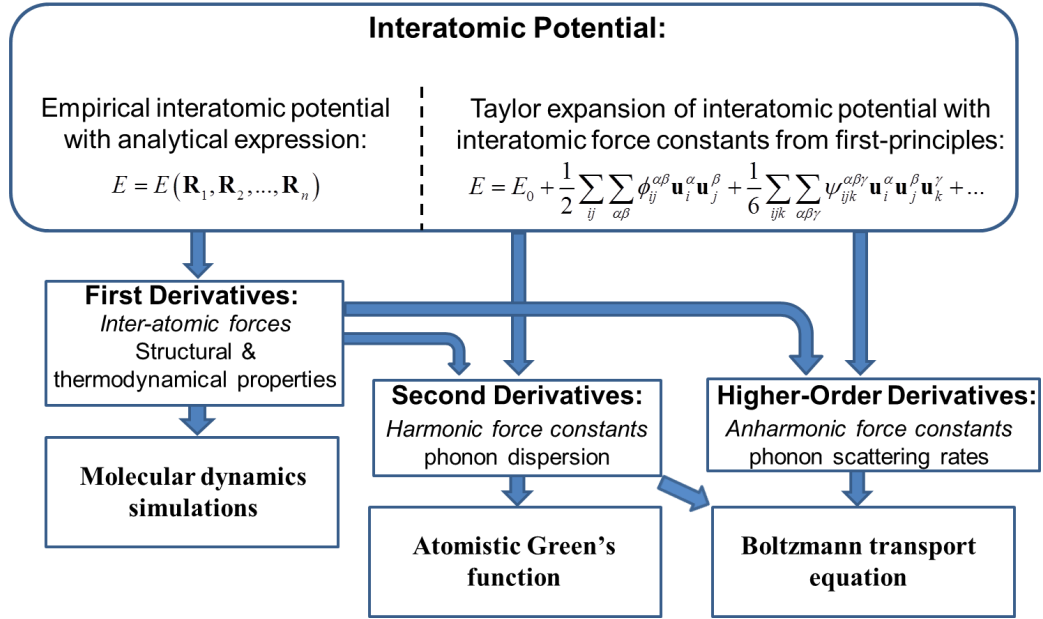

FIG. 6: Atomistic simulation methods for the calculation of phonon properties and thermal conductivity.

### 3.1 Interatomic potential and force constants

The key to the modeling of phonon properties and thermal conductivity of a nonmetallic material is its interatomic interaction. The interatomic potential of material, $E$, is a function of positions $\mathbf{R}_i$ of atoms:

$$E = E(\mathbf{R}_1, \mathbf{R}_2, ..., \mathbf{R}_n) \tag{5}$$

In the classical MD simulations, empirical interatomic potentials with simplified analytical expressions are used for Eq. (5). The expressions of empirical interatomic potentials are usually obtained by fitting only a few physical properties of the materials, such as lattice constant, elastic constants or interatomic forces, which make it challenging to reproduce phonon properties well and thus the accurate prediction of thermophysical properties including thermal conductivity. Though inaccurate, empirical interatomic potentials still gain considerable attention for some materials, such as diamond and silicon, due to their computationally efficiency. Many efforts are still devoted to develop empirical interatomic potentials for novel and not well-studied materials, including many 2D materials [49, 50].

To obtain phonon properties (phonon dispersion and phonon scattering rate), it is more convenient to re-write the interatomic potential of a crystal $E$ as the Taylor series with respect to atomic displacement from their equilibrium positions,

$$E = E_0 + \frac{1}{2}\sum_{\mathbf{R}\tau\alpha}\sum_{\mathbf{R}'\tau'\beta} \phi^{\alpha\beta}_{\mathbf{R}\tau,\mathbf{R}'\tau'} u^{\alpha}_{\mathbf{R}\tau} u^{\beta}_{\mathbf{R}'\tau'} + \frac{1}{6}\sum_{\mathbf{R}\tau\alpha}\sum_{\mathbf{R}'\tau'\beta}\sum_{\mathbf{R}''\tau''\gamma} \psi^{\alpha\beta\gamma}_{\mathbf{R}\tau,\mathbf{R}'\tau',\mathbf{R}''\tau''} u^{\alpha}_{\mathbf{R}\tau} u^{\beta}_{\mathbf{R}'\tau'} u^{\gamma}_{\mathbf{R}''\tau''} + ... \tag{6}$$

where $(\mathbf{R}, \tau, \alpha)$ standard for the degree of freedom that corresponds to the $\alpha$ direction of the $\tau$-th atom in the unit cell $\mathbf{R}$, $E_0$ is the total potential energy of the crystal when atoms stay in their equilibrium positions, $u$ is the displacement, and $\phi$ and $\psi$ are the second-order and third-order interatomic force constants.

Both empirical interatomic potentials and the first-principles calculations can be employed to extract the interatomic force constants in Eq. (6) to calculate phonon properties. The interatomic force constants can be obtained through either density-functional perturbation theory [51] or real-space small-displacement methods [52, 53] from first-principles calculations.

When the third-order and higher-order terms are neglected, known as the harmonic approximation, the relation between the phonon wavevector $\mathbf{q}$ and phonon frequency $\omega$, also known as phonon dispersion relation, is determined by the dynamical matrix $\mathbf{D}$, which is the function of second-order force constants $\phi$, i.e.

$$D_{\tau\tau'}^{\alpha\beta}(\mathbf{q}) = \frac{1}{\sqrt{m_\tau m_{\tau'}}} \sum_{\mathbf{R'}} \phi_{0\tau,\mathbf{R'}\tau'}^{\alpha\beta} e^{i\mathbf{q}\cdot\mathbf{R'}}, \tag{7}$$

where $m$ is atomic mass. The phonon frequency of the $s$-th phonon mode at $\mathbf{q}$ is the square roots of the $s$-th eigenvalue of the dynamical matrix $D_{\tau\tau'}^{\alpha\beta}(\mathbf{q})$. After obtaining the phonon dispersion the group velocity for phonon mode $\omega_{\mathbf{q}s}$ is given by $\mathbf{v}_{\mathbf{q}s} = \partial \omega_{\mathbf{q}s} / \partial \mathbf{q}$, and its heat capacity is $c_{\mathbf{q}s} = \hbar^2 \omega_{\mathbf{q}s} n_{\mathbf{q}s}^0 (n_{\mathbf{q}s}^0 + 1) / k_B T^2$, with the Planck constant $\hbar$, the Boltzmann constant $k_B$ and the temperature $T$. Here, $n_{\mathbf{q}s}^0$ is the Bose-Einstein distribution for phonons at equilibrium. .

Figure 7 shows the phonon dispersion of graphene and MoS$_2$ along high-symmetry directions, which are calculated according to Eq. (7) using the second-order (harmonic) force constants from the first-principles calculations [54, 55]. There are three acoustic phonon branches, i.e. longitudinal acoustic (LA), transverse acoustic (TA) and flexural acoustic branches (ZA), and three optical branches, including one longitudinal optical (LO), transverse optical (TO) and flexural optical (ZO) branches for graphene, as shown in Figure 6(a). For MoS$_2$, there are three acoustic branches and six optical branches, as shown in Figure 6(b) due to the three-atom basis in the unit cell. For one-atom-thick 2D crystals, such as graphene, the ZA and ZO branches correspond to pure out-of-plane movements of atoms. In contrast, the vibrations of ZA and ZO modes in non-one-atom-thick 2D crystal, such as MoS$_2$, involve both out-of-plane and in-plane movements of atoms, except at the zone center, where the vibration is purely out-of-plane.

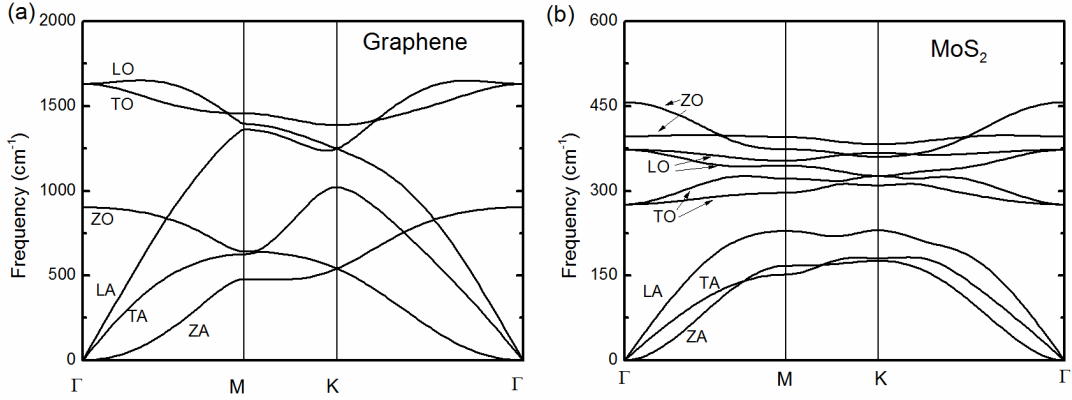

FIG. 7: Phonon dispersion for (a) graphene [54] and (b) MoS$_2$ [55] from the first-principles calculations.

While under the harmonic approximation phonons do not scatter with each other which would have led to infinite thermal conductivity, phonon-phonon scattering due to anharmonic terms in Eq. (6) is the key to the finite thermal conductivity. The phonon-phonon scattering rate can be estimated from Fermi's golden rule. The phonon-phonon scattering rate can be estimated from Fermi's golden rule. The total scattering rate of each phonon mode due to three-phonon processes is expressed as [56]

$$\Gamma_{\mathbf{q}s} = \frac{\pi}{2} \sum_{\mathbf{q}',\mathbf{q}'',s',s''} |V_3(-\mathbf{q}s,\mathbf{q}'s',\mathbf{q}''s'')|^2 \times \left[ \frac{1}{2}(1+n^0_{\mathbf{q}'s'}+n^0_{\mathbf{q}''s''})\delta(\omega_{\mathbf{q}'s'}+\omega_{\mathbf{q}''s''}-\omega_{-\mathbf{q}s}) \\ +(n^0_{\mathbf{q}'s'}-n^0_{\mathbf{q}''s''})\delta(\omega_{\mathbf{q}''s''}-\omega_{\mathbf{q}''s''}-\omega_{-\mathbf{q}s}) \right], \quad (8)$$

where $V_3(-\mathbf{q}s,\mathbf{q}'s',\mathbf{q}''s'')$ are the three-phonon scattering matrix elements, which is related to the third-order force constants. Figure 8 shows the calculated phonon lifetime $\tau_{\mathbf{q}s}\left(=1/\Gamma_{\mathbf{q}s}\right)$ of MoS$_2$ using the third-order force constants from first-principles.

With phonon dispersion and scattering rate calculated, the thermal conductivity of a material can be calculated using the simple kinetic theory [57], where $\kappa = \sum_{\mathbf{q}s} c_{\mathbf{q}s} v^2_{\mathbf{q}s} \tau_{\mathbf{q}s}$, or using the PBTE calculations, as will be discussed in Sec. 3.2.

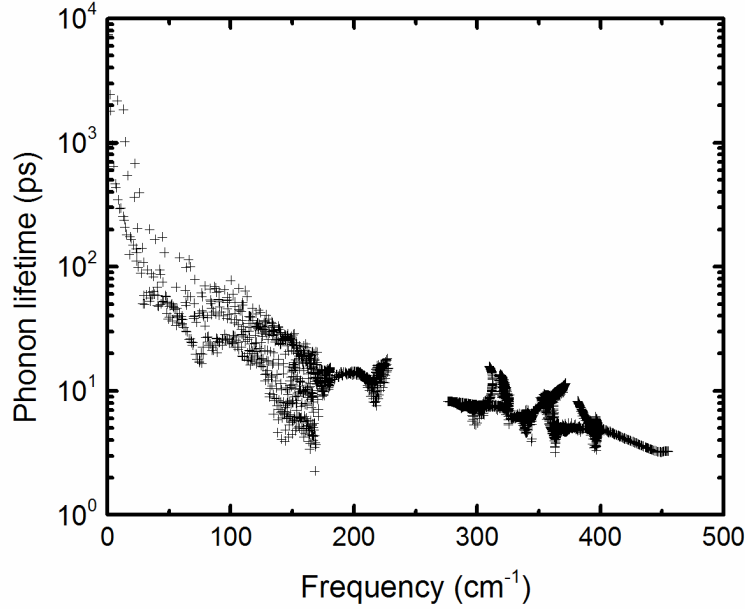

FIG. 8: Phonon lifetime in MoS$_2$ [55] from first-principles calculations.

The phonon dispersion and thermal conductivity of a few crystal materials computed using the interatomic force constants from first-principles as inputs agree well with experiment measurements, making the thermal conductivity calculation from the PBTE-based approach of predictive power. The atomic simulations integrated with first-principles calculations have envolved to be promising modeling tools to predict the phonon properties in crystal materials and to design novel materials with exceptional thermal properties.

## 3.2 Peierls-Boltzmann transport equation (PBTE)

Phenomenologically, the thermal conductivity, $\kappa$, of a material can be evaluated using Fourier's law of heat conduction, which relates a temperature gradient $\nabla T$ in the material to an applied heat flux $\mathbf{J}$ through

$$\mathbf{J} = -\kappa \cdot \nabla T. \qquad (9)$$

Strictly, the thermal conductivity in most materials are anisotropic and should be a tensor $\boldsymbol{\kappa}$. If the temperature gradient is along certain direction, say in the $x$ direction, the expression is simplified to

$$J^x = -\kappa^{xx} \cdot (dT/dx). \qquad (10)$$

The heat flux in a 2D crystal can be expressed as the summation of the heat carried by each phonon mode through

$$J^x = \frac{1}{(2\pi)^2 t} \sum_s \int \hbar \omega_{\mathbf{q}s} v^x_{\mathbf{q}s} n_{\mathbf{q}s} d\mathbf{q}, \qquad (11)$$

where the phonon distribution function $n_{\mathbf{q}s}(\mathbf{r},t)$ describes the occupation number of phonon mode $\mathbf{q}s$ at position $\mathbf{r}$ and time $t$. The phonon distribution function deviates from the equilibrium distribution $n^0_{\mathbf{q}s}(\mathbf{r},t)$ to $n_{\mathbf{q}s}(\mathbf{r},t)$ when there is a temperature gradient. The thermal conductivity can then be evaluated by combining Eq. (10) and Eq. (11)

$$\kappa^{xx} = -\frac{1}{(2\pi)^2 t(dT/dx)} \sum_s \int \hbar \omega_{\mathbf{q}s} v^x_{\mathbf{q}s} n_{\mathbf{q}s} d\mathbf{q}. \qquad (12)$$

As the frequencies and group velocities of phonon modes are given by phonon dispersion relation of the material, the thermal conductivity can be determined by solving PBTE for the phonon distribution function, which considers the balance between phonon diffusion driven by a small temperature gradient and phonon scattering. In the steady state and under small temperature gradient, the PBTE could be written in the linearized form as [56]

$$-\mathbf{v}_{\mathbf{q}s} \cdot \nabla T \frac{\partial n^0_{\mathbf{q}s}(\mathbf{r})}{\partial T} = \left.\frac{\partial n_{\mathbf{q}s}(\mathbf{r})}{\partial t}\right|_{scatt}. \qquad (13),$$

The scattering mechanisms of phonons include phonon-phonon scattering, phonon-boundary scattering, phonon-defect scattering and so on. The main scattering to drive the temperature to the equilibrium in a large defect-free crystal is phonon-phonon scatterings. Phonon-phonon scattering couples the phonon distribution function for mode $\mathbf{q}s$ with other phonon modes, making a great challenge to solve PBTE for phonon distribution function. Due to the coupling among different phonon modes which requires both energy and momentum conservation and complexity of the scattering terms, PBTE is always solved under some approximations. The most commonly used one is the so-called single-mode relaxation time approximation (SMRTA), where the scattering of certain phonon mode is irrelevant to the condition of other phonon modes, but other phonon modes are assumed to stay in their equilibrium condition. Under such an approximation, the state of each phonon mode can be solved independently, and Eq. (13) reduces to

$$-\mathbf{v}_{\mathbf{q}s} \cdot \nabla T \frac{\partial n^0_{\mathbf{q}s}(\mathbf{r})}{\partial T} = \frac{n_{\mathbf{q}s}(\mathbf{r}) - n^0_{\mathbf{q}s}(\mathbf{r})}{\tau_{\mathbf{q}s}}, \qquad (14)$$

where $\tau_{\mathbf{q}s}$ is the relaxation time, or inverse of scattering rate $1/\Gamma_{\mathbf{q}s}$, of mode $\mathbf{q}s$. Assuming the individual scattering event, whose relaxation time (inverse to the scattering rate) is $\tau^i_{\mathbf{q}s}$ ($1/\Gamma^i_{\mathbf{q}s}$), is independent to others,

$$1/\tau_{\mathbf{q}s} = \sum_i 1/\tau^i_{\mathbf{q}s} \quad (\Gamma_{\mathbf{q}s} = \sum_i \Gamma^i_{\mathbf{q}s}). \qquad (15)$$

The thermal conductivity of 2D materials under relaxation time approximation is then expressed as

$$\kappa^{xx} = \frac{\hbar^2}{(2\pi)^2 k_B T^2 t} \sum_s \int \omega^2_{\mathbf{q}s} (v^x_{\mathbf{q}s})^2 n^0_{\mathbf{q}s}(n^0_{\mathbf{q}s}+1)\tau_{\mathbf{q}s} d\mathbf{q}. \qquad (16)$$

For bulk material, there exist a variety of analytical and empirical models for the relaxation times due to different scattering mechanisms in Eq. (15), which leads to reasonable agreement of theoretical prediction and experimental measurement on the thermal conductivity of bulk materials. Due to the simplicity of the analytical form, some researchers have applied this analytical expression to estimate the thermal conductivity of 2D materials under the framework of relaxation time approximation [58, 59]. To model the phonon-phonon scatterings, usually only Umklapp processes are considered, whose scattering rates or scattering matrix elements are given by the expressions derived by Klemens [60, 61]. However, the Klemens expressions for matrix elements for three-phonon scatterings are derived for an isotropic solid, assuming a linear dispersion of phonon modes and without any details of phonon eigenvectors. Therefore, the validity of using such expressions to estimate the thermal conductivity is not confirmed for 2D materials. The relaxation time due to phonon-phonon scattering can also be estimated by phonon lifetime, whose expression is given in Eq. (8). Such approximation works reasonably well for materials with strong Umklapp scattering, but always severely underestimates thermal conductivity when the normal scattering is strong [55, 62].

Recently, Broido *et al*. [44] extended Omini and Sparavigna' work [63] to estimate the phonon scattering rates of any three-phonon processes using Fermi's golden rule and to strictly solve the PBTE through iterative approach. In particular, when the interatomic force constants from first-principles are incorporated in the calculation, the calculated thermal conductivities of a wide range of materials agree well with the experimental measurement [45, 64, 65]. The success of this approach on thermal conductivity prediction has stimulated quite a few first-principles simulations on the phonon properties and thermal conductivity in 2D crystals [54, 55, 66, 67].

In the following tables summarizing the computational studies, we will use PBTE-iterative, PBTE-SMRTA-LT and PBTE-SMRTA-Klemens to refer to the approaches that solve PBTE with iterative approach, solve PBTE with SMRTA and using phonon lifetime as the relaxation time, and solve PBTE with SMRTA and using the relaxation time calculated from Klemens expressions.

### 3.3 Molecular dynamics simulations

In MD simulations, the movements of atoms within a small computational domain are simulated and recorded according to Newton's second law. Both non-equilibrium molecular dynamics (NEMD) [68] and equilibrium molecular dynamics (EMD) [69] have been developed to calculate thermal conductivity of bulk, nanostructured and 2D materials. In NEMD, heat flux is applied by inserting/removing energy into/out of the system. When the steady temperature gradient is established along the heat flux direction, the thermal conductivity can be calculated using Fourier's law of heat conduction. In EMD, heat current in $\alpha$ direction $S^\alpha(t)$ computed, which is expressed as

$$S^\alpha(t) = d\left(\sum_i \mathbf{r}_i h_i\right)/dt \tag{17}$$

where $h_i$ is the total energy of particle *i*. By averaging the heat-current autocorrelation function $\langle S^\alpha(t)S^\alpha(0)\rangle$, the thermal conductivity is then calculated using the Green-Kubo formula,

$$\kappa^{\alpha\alpha} = \frac{1}{k_B V T^2} \int \langle S^\alpha(t) S^\alpha(0) \rangle dt. \tag{18}$$

MD simulations can also provide the detailed mode-dependent information of phonon, including phonon dispersion through velocity-autocorrelation function [70] and phonon lifetime using the normal mode analysis [71]. The interaction of phonon waves with interfaces and defects have also been simulated through the wave-packet method [72].

One great advantage of MD simulations is that the dynamics of the materials is studied in the real space, where a variety of physical effects such as strain, doping, isotope, defects and the size of sample can be easily studied. However, there are some challenges for MD simulations. First, MD simulation is a classical method, where quantum effects are not considered. Phonons in MD simulation follow the Maxwell-Boltzmann distribution, not the Bose-Einstein one. Although quantum correction for heat capacity has been proposed, the phonon scattering rates which are related to the phonon distribution function are still governed by the classical theory. This has limited the applicability of MD simulation for temperature far below the Debye temperature [73]. Second, MD simulations highly depend on interatomic potentials used. Since the empirical potentials are usually developed by fitting a limited set of material properties, they might not give a satisfactory description on other properties that are not included in the database for potential development. Unfortunately, phonon transport properties, including phonon dispersion, and higher-order anharmonic properties, e.g. Gruneisen parameters and thermal expansion coefficients, are seldom included in the database for potential development, except some recent work [74]. Recently, some efforts are exerted to construct interatomic potential with interatomic force constants from first-principles calculations, which could improve the accuracy of the potential significantly [45, 75]. Third, MD simulations are carried out in real space, which makes the simulation time and computational domain finite leading to systemic error on the calculated thermal conductivity. To avoid the simulation-time-dependent results, the simulation time has to be much longer than the longest phonon relaxation time of the phonon modes existing in the simulation system [45]. To eliminate the size effects due to limited simulation domain, some techniques have been proposed for EMD [76] and NEMD [43], but this requires independent calculations with different computation domain sizes.

MD simulations have been applied to predict the thermal conductivities of many 2D materials, including graphene, graphene derivatives, silicene and $MoS_2$, and nanostructures of 2D crystals. Some of these results will be reviewed in Sec. IV.

### 3.4 Atomistic Green's function (AGF)

While anharmonic phonon-phonon scatterings dominate the thermal resistance for a perfect bulk crystalline material at high temperature, the elastic scatterings, such as boundary scatterings and defect scatterings, become important at low temperature. For 2D materials, defects and grain boundaries are very likely to occur. It is therefore highly desirable to study the elastic scatterings in 2D materials. When the anharmonic scattering is not as important, the vibration of an atomic system is modeled by a harmonic spring system. The AGF approach was proposed to solve the dynamical equation of the harmonic system [46, 77, 78], which gives phonon transmission function $Tr(\omega)$, the probability that phonons with given frequency emit from one thermal reservoir and eventually reach the other thermal reservoir. The thermal conductance can then be calculated with the Landauer formalism using the total phonon transmission [79]:

$$G = \frac{1}{2\pi A} \int_0^\infty \hbar\omega \frac{\partial n^0(\omega,T)}{\partial T} Tr(\omega) \text{DOS}(\omega) d\omega, \qquad (19)$$

where $A$ is the cross-sectional area perpendicular to the heat flow direction, DOS is the phonon density of states for given frequency. Given the length of the sample, the thermal conductivity is calculated as $\kappa = GL/A$, with the sample length $L$.

Many studies had employed the AGF to calculate the thermal conductivity of 2D materials with boundary scatterings (nanoribbons) and defect scatterings. Strictly, the calculated thermal conductivity from the AGF calculations is only valid at very low temperature or for very short samples. However, by exploring how the transmission coefficient is influenced by the imperfectness of the lattice, the role of these factors on thermal conductivity can be explored even at high temperature. In particular, recent development on converting the transmission function to scattering cross section might make possible to consider both anharmonic phonon-phonon scatterings and elastic scatterings in the PBTE framework [80-82].

## 4 Thermal transport in some typical 2D materials

In this section, we summarize the available experimental data and theoretical calculations on the thermal conductivity of a few typical 2D materials, including graphene, h-BN, silicene, TMDs, bismuth telluride and black phosphorus.

### 4.1 Graphene

Many excellent reviews on thermal transport in graphene have been published in the past few years [17-20]. Here we just list a few important works. Most of results are summarized in Fig. 9, and Table I and II.

As shown in Fig. 9, the reported values of measured thermal conductivity for single-layer graphene is scattered, ranging from 2000 W/mK to 5500 W/mK at room temperature, owing to both sample quality and experimental uncertainty. The thermal conductivity reduces to around 600 W/mK at 680K. Despite the large uncertainty, the thermal conductivity in graphene is larger than other carbon allotropes, such as graphite and diamond, which renders great excitements on the applications of graphene as heat spreaders for electronics and optoelectronics.

The first measurement on the thermal conductivity of graphene was reported by Balandin *et al.* [25] using optothermal Raman technique as described in Section 2.1. In their experiment, a single-layer graphene was suspended over a 3-μm-wide trench and heated by a 488 nm wavelength laser with spot size of about 0.5-1.0 μm. The temperature response of G peak of the Raman signal was used as thermometer. To obtain the optical absorption coefficient, the authors measured the Raman intensity of highly-oriented bulk pyrolytic graphite (POG) samples, and then converted it to the optical absorption of single-layer graphene. They calculated the power absorbed by graphene to be about 13% of the laser power. By assuming that the supported graphene has the same thermal conductivity as the suspended graphene and that the graphite covered on the top of the supported graphene serves as a perfect heat sink, they obtained the thermal conductivity to be 5300 W/mK at room temperature, a value that is two times higher value than HOPG and diamond.

Later, in Ghosh *et al.*'s experiments [83], the optical absorption was refined to 11-12% when taking into account the absorption of the rough silicon underneath the trench, which yields a thermal conductivity to be 4100-4800 W/mK. Ghosh *et al.* [27] also showed that the thermal conductivity of few-layer graphene decreases with the number of graphene monolayers. Faugeras *et al.* [32] and Lee *et al.* [31] employed a much lower optical absorption of 2.3%, which was confirmed by optical reflectivity and transmission measurements [84-86] as well as theoretical calculation based on the general noninteracting tight-binding model [87]. They found a much lower thermal conductivity than those from Balandin *et al.*'s [25] experiments. However, as pointed by Nika and Balandin [20], the optical absorption of 2.3% is observed only in the near-infrared wavelength and the optical absorption of graphene increases with the wavelength of laser due to many-body effects. Cai *et al.* [30] directly measured the optical absorption by placing a detector under the graphene sheet and found a 3.3% optical absorption of graphene when a 532 nm laser beam is used. In addition, they measured the thermal conductivity of supported graphene $\kappa_{sup}$ and interfacial thermal conductance $g$ between graphene and substrate using independent Raman measurements. By considering contact resistance and the thermal conductivity of the supported graphene, which is different from the suspended one, in the heat conduction model, they reported a thermal conductivity about (2500+1100/-1050) W/mK near 350 K for the suspended graphene. Chen *et al.* [33] also measured the thermal conductivity of graphene in ambient condition and in vacuum, and they found a measurable difference on the extracted thermal conductivity due to the heat loss to the air. Pettes *et al.* [26] developed a graphene transfer process to assemble suspended CVD graphene on micro-bridge devices and measured the thermal conductivity of bi-layer graphene sheets to be 600 W/mK. They attributed the low thermal conductivity they obtained to a residual polymeric layer due to their transfer process. Xu *et al.* [88] explored the length-dependent thermal transport in monolayer graphene using micro-bridge methods and observed a logarithmic dependence on sample length.

Many theoretical calculations have been conducted after the first exciting measurement of high thermal conductivity in graphene. Nika *et al.* [89] used the relaxation time approximation, in which the three-phonon scattering matrix elements are calculated from the expression derived by Klemens [60]. In their work, the flexural acoustic phonons are neglected due to their small group velocity, as indicated in Fig. 3(a). However, Lindsay *et al.* [62] employed the PBTE formalism to calculate the thermal conductivity of graphene with the interatomic force constants from modified Tersoff potentials, and found that the flexural acoustic phonons conducted 70-80% heat. This was confirmed by Singh [90] with a similar approach and later by several first-principles-based PBTE simulations [54, 66, 91]. The large contribution of flexural acoustic branch is due to the symmetry selection rule [62], where any three-phonon processes involving odd number of flexural phonon modes (ZA and ZO) cannot happen, which results in much larger relaxation time for the flexural acoustic phonon modes. Apart from the low scattering rates of these phonons modes, the scattering of long-wavelength flexural acoustic phonons are usually through normal scatterings, which do not directly contribute to thermal resistance. When a heat flux is imposed, a large portion of phonons from other branches are converted to the flexural acoustic phonons, which in turn enhances the contribution from acoustic flexural modes. This can be confirmed by comparing the contribution of ZA phonons from SMRTA [92] and iterative solution [54]. It was found that the contribution from ZA modes increases from 50% to 80% when changing SMRTA solution to iterative solution. Due to the conversion from other phonon modes to ZA modes, the thermal conductivity is also increased from 500-550 W/mK to around 3000 W/mK.

Many MD simulations have been conducted to predict the thermal conductivity of graphene. However, the data is rather scattered due to the inaccuracy of the potentials used. Although not reliable on the absolute values, MD simulations provide great qualitative trends on the effects of size, defects, strain and the interaction with surroundings on thermal conductivity of 2D materials since MD simulations are on real space where these effects can be easily incorporated. We will discuss these effects in Section 5-7.

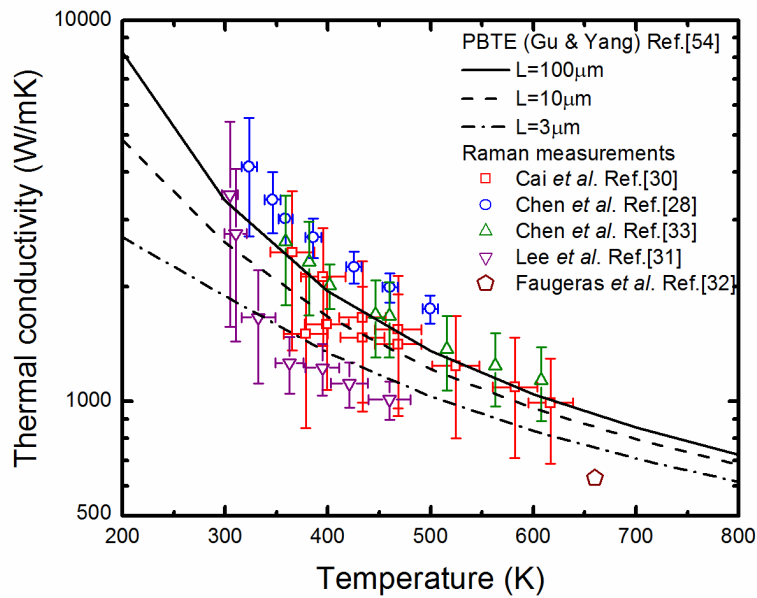

FIG. 9: Measured temperature-dependent thermal conductivity of single-layer graphene. The thermal conductivity of the samples with naturally-occurring C isotopes were reported by Cai *et al.* [30], Chen *et al.* [33], Lee *et al.* [31] and Faugeras *et al.* [32]. The thermal conductivity of isotopically-pure $^{12}$C (0.01% $^{13}$C) graphene was reported by Chen *et al.* [28], The calculated thermal conductivity of graphene with different sample length from PBTE by Gu and Yang [54] is also plotted.

## 4.2 Boron nitride

Single-layer hexagonal boron nitride (h-BN) has a similar lattice structure as graphene, but with carbon atoms replaced by alternating B and N atoms, as shown in Fig. 1(b). With a wide band gap ~5.8 eV, h-BN has been regarded as the most promising candidate for the dielectric layer in graphene electronics. h-BN substrate has been reported to improves carrier mobilities of graphene much better than that with $SiO_2$ substrate [109], because it has an atomically-smooth surface that is relatively free of dangling bonds and charge traps. Unlike graphene, limited work has been published on the thermal transport properties of 2D h-BN sheet. Available experimental measurements and theoretical calculations are presented in Fig. 10 and Table III.

The thermal conductivity of a single-layer h-BN was studied using EMD with Tersoff potential parameterized by Sevik *et al*. [110] and by iterative PBTE method using Tersoff potential with another set of coefficients developed by Lindsay *et al* [111, 112]. The calculated thermal conductivity from EMD is 400 W/mK at 300 K [110], while the work using PBTE reported a thermal conductivity of 600 W/mK for 2-μm-long h-BN sheet and found a strong length-dependence on the thermal conductivity [111]. Lindsay *et al.* [111] also studied the thermal transport in single-layer h-BN with naturally occurring B and N isotopes and found that the thermal conductivity of a 10-μm-long h-BN sheet can be enhanced by 37% by isotopic enrichment because of the large concentration of $^{10}B$ atoms (19.9%) in the naturally occurring h-BN. They also studied the thermal transport in few-layer h-BN and observed a monotonic decrease of thermal conductivity, which approaches to bulk value as the increase of layer number [112].

Jo *et al.* [39] measured the thermal conductivity of suspended few-layer h-BN using a micro-bridge device. The room-temperature thermal conductivity of an 11-layer sample was found to be about 360 W/mK, which is close to the measured basal-plane thermal conductivity of bulk h-BN in literature [113]. As the thickness of the sample reduced to 5 layers, the thermal conductivity is reduced to 250 W/mK. The reduction of thermal conductivity in 5-layer sample becomes more significant in low temperature. The experimental data is contradictory to Lindsay *et al.*'s [112] calculation which shows that the thermal conductivity of few-layer h-BN decreases as the layer number increases. They attributed this difference to the presence of a polymer residue layer on the sample surface since the polymer residues provide additional scatterings to low-frequency phonons, especially for thinner samples. In a recent Raman measurement, Zhou *et al.* [114] measured the temperature-coefficients of the 2G band for monolayer, bilayer and 9-layer h-BN. After transfer the h-BN sample to glass slides, they measured the optical absorption, and obtained a 5% optical absorption for 9-layer h-BN. The extracted thermal conductivity of 9-layer h-BN is 227-280 W/mK at room temperature, which is close to Jo *et al.*'s data for 5-layer h-BN.

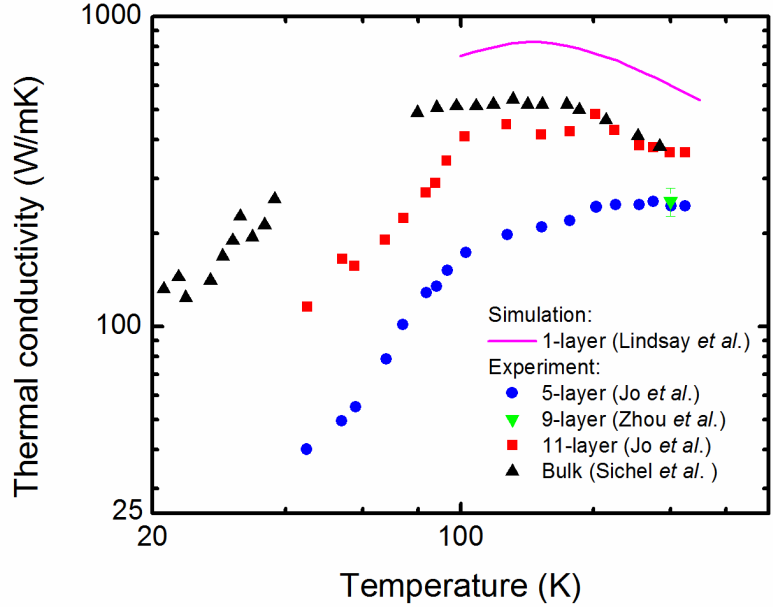

FIG. 10: Measured temperature-dependent thermal conductivity of h-BN, as reported by Jo *et al.* [39] for 5-layer and 11-layer samples, Zhou *et al.* [114] for 9-layer sample, and Sichel *et al.* [113] for bulk sample. The calculated thermal conductivity of 2-μm-long single-layer h-BN from PBTE by Lindsay *et al.* [111] is also plotted.

### 4.3 Silicene

Silicene is a two-dimensional graphene-like honeycomb silicon crystal, which was predicted to be stable theoretically in early 1990s [117] and has been synthesized recently [118, 119]. Silicene might be more easily integrated with silicon-based semiconductor devices. Due to its similarities with the lattice structure of graphene, silicene shares many similar electronic properties with graphene. For example, the charge carrier of silicene is massless fermion just like those in graphene [119, 120]. However, compared with the planar structure of graphene, the honeycomb lattice of silicene is slightly buckled, which leads to some new exciting characteristics. For instance, the buckled structure breaks the symmetry of the crystal, making it possible to open a bandgap by applying electric field, which is a nontrivial challenge for graphene [121, 122].

Similarly, it is of great interest to explore the role of buckled lattice on phonon transport mechanisms in silicene and compare that with the planar graphene. The thermal conductivity of silicene has been investigated by first-principles-based PBTE approach and classical MD simulations, as summarized in Table IV.

Gu and Yang calculated the thermal conductivity of single-layer silicene using a first-principles-based PBTE approach [54]. The calculated temperature dependence of thermal conductivity is shown in Fig. 11. This is dramatically different from graphene, whose thermal conductivity is comparable to or even larger than its bulk forms, graphite or diamond. The thermal conductivity of silicene is unexpectedly an order of magnitude lower than that of bulk silicon. For example, at 300K, the thermal conductivity of a silicene sheet with $L$=10 μm is only 26 W/mK compared with ~140 W/mK for bulk silicon. Indeed, the thermal conductivity of single layer silicene with $L$=10μm is comparable to the thermal conductivity of silicon nanowires with a diameter of 55 nm and thin films with a thickness of 20 nm. When sample length decreases, the thermal conductivity is further reduced and exhibits weaker temperature dependence due to the stronger phonon-boundary scattering. Further examination of the contribution of thermal conductivity from different phonon branches for 10-μm-long silicene shows that in silicene only around 7.5% of heat is conducted by the ZA branch while the ZA branch contributes about 75% of the large thermal conductivity of graphene. The LA and TA branches together contribute 20% and 70% to the total thermal conductivity for graphene and silicene, respectively. The contribution of the ZA branch is related to the breakdown of the selection rule of symmetry, as discussed in Sec. 4.1. In addition, they observed a length-dependent thermal conductivity even when the length of the silicene sample is larger than 30 μm. Such strong length-dependence is attributed to the linear dispersion of ZA branch near the Γ point (**q**=0), which could be universal for non-atom-thick 2D materials with hexagonal lattice. The detailed discussions on length-dependent phonon transport are presented in Sec. 4.1. Xie *et al.* also performed similar first-principles simulations, but they found a much smaller thermal conductivity [123]. In addition, the relaxation times of long wave-length acoustic phonon modes from their work approach zero as the wavevector $q \to 0$, which is opposite to the scaling relation that Gu and Yang [54] derived in their work, and also contradicts with the classical theory that low-frequency phonon modes are not likely to be scattered. It is desirable to have more detailed analysis to clarify the difference between these theoretical works.

Many MD simulations have been performed with empirical interatomic potentials including the Tersoff potential [124], modified embedded-atom method (MEAM) potential [125] and the modified Stillinger-Weber potential [49]. The reported thermal conductivity from MD simulations ranges from 5 to 50 W/mK. The large difference on the thermal conductivity value can be attributed to the inaccuracy of empirical potentials and thus phonon dispersion and phonon scattering rates, as well as the simulation parameters employed. We note that the ranges of phonon spectra from the Tersoff potential and the modified Stillinger-Weber potential are reasonably close to that from first-principles calculations. However, the Tersoff potential and the original Stillinger-Weber potential cannot reproduce the buckled lattice. Considering the important role of the structure buckling on thermal transport, which leads to additional scattering compared with the planar structure, it might not be a good choice to employ these two potentials for silicene, although they have been successfully applied to bulk silicon. Instead, the silicene crystal structure predicted by the modified Stillinger-Weber and MEAM potentials is buckled with a 0.42 Å [49] and 0.85 Å [126] buckling distance, respectively. Therefore, these two potentials are more suitable to be used for qualitatively investigating the mechanisms of some geometric and physical factors on the phonon transport in silicene, which are relatively challenging to access using the first-principles-based PBTE approach. The effect of strain, defects, isotopes, and the size of nanoribbons on the thermal conductivity of silicon have also been explored using MD simulations, as listed in Table IV.

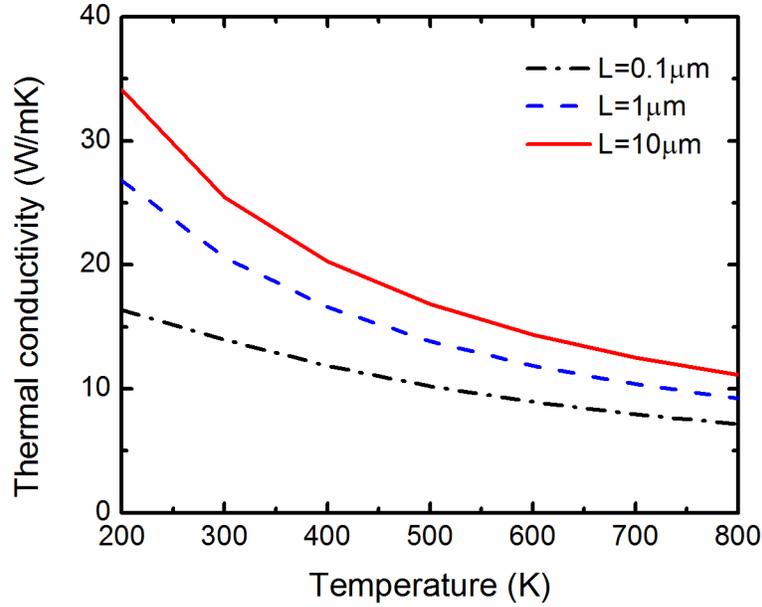

FIG. 11: Calculated temperature-dependent thermal conductivity of silicene from fist-principles-based PBTE calculation [54].

## 4.4 Transition metal dichalcogenides (TMDs)

Monolayer TMD have a three-layer structure in which one layer of transition metal atoms are sandwiched by two layers of chalcogenide atoms (S, Se, Te, ….). Depending on how the chalcogenide atoms sit on each side of the metal layer, there are two polymorphs for monolayer TMDs: 1T phase with $D_{3d}$ point group and 2H phase with $D_{3h}$ point group, as shown in Fig.1(e,f). Some of the interesting physical and chemical properties of single-layer TMDs have been explored and are reviewed in Ref. [7, 8], with heightened interests in technological applications of TMD. For example, $MoS_2$ field-effect transistors can be effectively switched off with a high on/off switching ratio exceeding $10^8$ at room temperature [131], which are difficult to achieve in graphene transistors. The success of fabricating photo-transistor made from single-layer $MoS_2$ also paves the way to the development of multifunctional optoelectronic devices using 2D TMDs [132, 133].

According to the conventional wisdom, the thermal conductivities of TMDs are thought to be low due to their heavy atom mass of the transition metals and the low Debye temperature. As a result, many studies have exploited the use of single-layer or few-layer TMDs as potential high efficiency thermoelectric materials. However, recently a few theoretical and experimental studies show that the thermal conductivity of $MoS_2$ is relatively high, as summarized in Fig. 12 and Table V.

Li *et al*. performed PBTE calculation on phonon transport in $MoS_2$ under the SMRTA, in which they studied the phonon linewidths of optical phonons and the thermal conductivity [67]. The calculated thermal conductivity of the isotopically-pure $MoS_2$ is 90 W/mK, while that of the $MoS_2$ with naturally-occurring Mo and S isotopes decreases to 71 W/mK. They identified the importance of the isotope scattering on the phonon linewidths, which is inversely proportional to the phonon relaxation time. Gu and Yang [55] predicted the thermal conductivity employing the iterative approach to solve the PBTE. They found that the normal process is important with large sample size. When the sample size is 10 μm, the thermal conductivity predicted using iterative solution is 20% higher than the prediction from the SMRTA. The contributions of different phonon branches to the total thermal conductivity is more like silicene than as graphene, since $MoS_2$ is not a graphene-like "one-atom-thick" 2D crystal.

MD simulations have been performed to explore the effect of layer numbers, strain and edge on the thermal conductivity of $MoS_2$ and $MoS_2$ nanoribbons. Several empirical potentials have been developed for $MoS_2$, including Brenner potential [134], Stillinger-Weber potential [50], and other force field models [135, 136]. The thermal conductivity values from MD simulations range widely from 1.3 W/mK to 44 W/mK, but mostly much lower than first-principles-based calculations.

Experimentally, Sahoo *et al*. [29] performed the first thermal conductivity measurement of few-layer $MoS_2$ (11 layers) through the optothermal Raman measurement. The obtained thermal conductivity of 11-layer $MoS_2$ is 52 W/mK at room temperature, which was extracted from the first-order temperature coefficient of $A_{1g}$ peak. Yan *et al*. [137] measured the exfoliated monolayer $MoS_2$ and found the thermal conductivity to be 34.5 W/mK. Jo *et al*. [138] used the micro-bridge device to measure the 4-layer and 7-layer $MoS_2$. Taube *et al*. [139] measured the supported monolayer $MoS_2$ on $SiO_2$/Si. Compared with Yan *et al*.'s suspended $MoS_2$, the supported monolayer is of a higher thermal conductivity value at 62 W/mK. The measured thermal conductivities for both single-layer and few-layer $MoS_2$ are considerably lower than the measured value of bulk $MoS_2$ from a pump-probe-based measurement, which is about 110 W/mK [140].

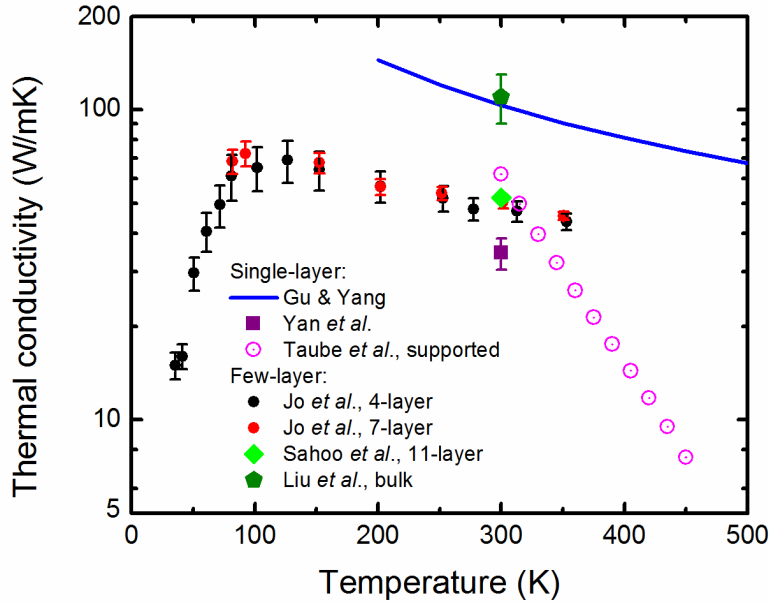

FIG. 12: Measured temperature-dependent thermal conductivity of MoS$_2$, as reported by Yan *et al*. [137] for single-layer sample, Jo *et al*. [138] for 4-layer and 7-layer samples, Sahoo *et al*. [29] for 11-layer sample, Liu *et al*. [140] for bulk sample, and Taube *et al*. [139] for supported single-layer sample. The calculated thermal conductivity of single-layer MoS$_2$ with a length of 1 μm from first-principles-based PBTE by Gu and Yang [55] is also plotted.

The studies are relatively rare for other transition metal dichalcogenides. In addition to MoS$_2$, Gu and Yang calculated the thermal conductivity of other TMDs, MX$_2$ (M = Mo, W, Zr, and Hf, X = S and Se) [55]. They found that the lattice thermal conductivities of 2H-type TMDs are above 50 W/mK at room temperature while the thermal conductivity values of the 1T-type TMDs are much lower. A very high thermal conductivity value of 142 W/mK was found in the single-layer WS$_2$. They attributed this large value to the large atomic weight difference between W and S, which leads to a very large phonon bandgap prohibiting the scattering between acoustic and optical phonon modes. Peimyoo *et al*. [141] measured the thermal conductivity of monolayer and bilayer CVD-grown WS$_2$ using the optothermal Raman method. The thermal conductivities at room temperature were estimated to be 32 and 53 W/mK for monolayer and bilayer WS$_2$, respectively.

### 4.5 Bi$_2$Te$_3$ quintuple

Bi$_2$Te$_3$ is also a *van der* Waals crystal with layered structure. Bi$_2$Te$_3$ is well recognized as one of the high-efficiency thermoelectric materials at room temperature with the very low thermal conductivity. Similar to other *van der* Waals crystals, it is also possible to mechanically exfoliate it to thin 2D sheets. Thin Bi$_2$Te$_3$ films down to a few quintuple layers or even Te-Bi, Bi-Te-Bi two-or-three-layer sheets have been fabricated using mechanical exfoliation [144]. By stacking the few-layer quintuples with different thickness on the top of each other, the pseudo-superlattice was also assembled. The in-plane thermal conductivity of the pseudo-superlattice are of the around 0.7~1.1 W/mK, which is a reduction by a factor of 2 from its bulk value of 1.7 W/mK [145, 146]. EMD has been conducted to study the layer-dependence of thermal conductivity of $Bi_2Te_3$ films. It was found that monolayer $Bi_2Te_3$ is of the highest thermal conductivity and the minimum thermal conductivity occurs when there are three quintuple layers, while the thermal conductivity of the bulk falls in between [147].

### 4.6 Black Phosphorus

Black phosphorus recently gained a lot of attention. A few groups performed first-principles-based PBTE calculations [148-150]. Although similar approaches were used, the thermal conductivity values reported vary from 30.15 to 110 W/mK along the zigzag direction and 13.65 to 36 W/mK along the armchair direction. The possible origins for the difference could be attributed to the choice of the cutoffs for interatomic force constants, the number of integration points in the first Brillouin zone used to calculate the phonon scattering rate and the thermal conductivity, and whether the iterative solution or the SMRTA of PBTE is employed to calculate the thermal conductivity.

Optothermal Raman measurement has also been applied to measure the thermal conductivity of few-layer black phosphorus [151]. The in-plane thermal conductivities along both zigzag direction and armchair direction exhibit strong thickness dependence. The thermal conductivities are ~40 W/mK along both the zigzag and the armchair directions for the thin film thicker than 15 nm, they are reduced to ~20 W/mK and ~10 W/mK as the film thickness is reduced, respectively.

## 5 Size-dependent thermal transport in 2D materials

Interfaces could induce additional scattering channels to phonons, leading to thermal conductivity reduction [57]. For 2D materials, the above simplified interface scattering picture might not necessarily hold. Depending on which dimension of a 2D crystal is tailored, as illustrated in Fig. 2(a), the effect of sample sizes could be quite different. In this section, we will discuss the length-dependent thermal transport when the size of the sample along the applied heat flux direction is changed, the thermal transport in nanoribbons where the size perpendicular to the direction of the applied heat flux is tuned, and the thickness-dependent transport where the layer number of the stacked *van der* Waal 2D crystal is changed. We will also explore whether the intrinsic thermal conductivity for an infinitely large 2D crystal is well defined.

### 5.1 Length-dependent thermal transport

Fourier's law of heat conduction can be strictly derived from the PBTE equation under the assumption that all heat carriers travel diffusively. If the feature size of sample is smaller than the mean free paths of some phonons, some of these phonons travel a long distance ballistically without scattering with other phonons until reaching the geometric boundaries.

Many studies reported length-dependent thermal transport in 2D materials [54, 66, 88, 91, 152], even when the sizes of samples are larger than 10 μm, which is far beyond a size of 800 nm, the averaged mean free path of graphene with the largest thermal conductivity at room temperature [83]. Figure 13 presents length-dependent thermal conductivity for suspended graphene from experimental measurements as well as atomistic simulations. In the experiment using the micro-bridge method [88], the measured thermal conductivity of graphene increases from 250 W/mK to 1700 W/mK when the sample length is increased from 300 nm to 9 μm, with a logarithmic dependence on the sample length. The length-dependent thermal conductivity was also observed for supported graphene. Considerable increase in thermal conductivity (from 280 to 580 W/mK) was reported for $SiO_2$-support graphene when its length increases from 260 nm to 10 μm [153].

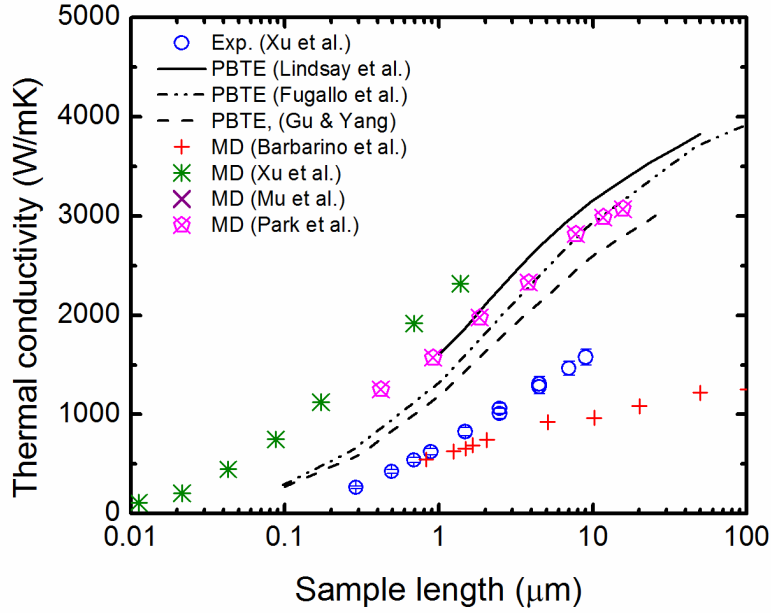

FIG. 13: Length-dependent thermal conductivity of graphene from micro-bridge measurement by Xu *et al*. [88], PBTE calculations by Lindsay *et al*. [66], Fugallo *et al*. [91], and Gu and Yang [54], MD simulations by Barbarino *et al*. [152], Xu *et al*. [88], Mu *et al*. [154] and Park *et al*. [155].

The size-dependent thermal conductivity of graphene has also been confirmed by first-principles-based PBTE calculations [54, 66, 91] and some NEMD simulations [88, 152, 154, 155], as shown in Figure 13. In the PBTE calculations, the effect of sample length is considered by using a boundary scattering term, which is written as

$$1/\tau_{\mathbf{q}s}^{B} = 2|v_{\mathbf{q}s}^{x}|/L \qquad (20)$$

where $L$ is the length of the 2D crystal, or the distance between two thermal reservoirs. In NEMD, the size of the sample is simply taken into account by simulating the thermal conduction in the samples with different sizes where the physical effect is sometime complicated by numerical artifacts as well. Both theoretical calculations and experiments showed that the thermal conductivity of graphene increase logarithmically with the increase of the sample size from 1um to 10 μm [88].

PBTE-based numerical calculations have also been carried out for other 2D materials aside from graphene, including h-BN [111], silicene [54] and MoS$_2$ [55]. The thermal conductivity also shows a logarithmic-like dependence on the sample size, as summarized in Figure 14.

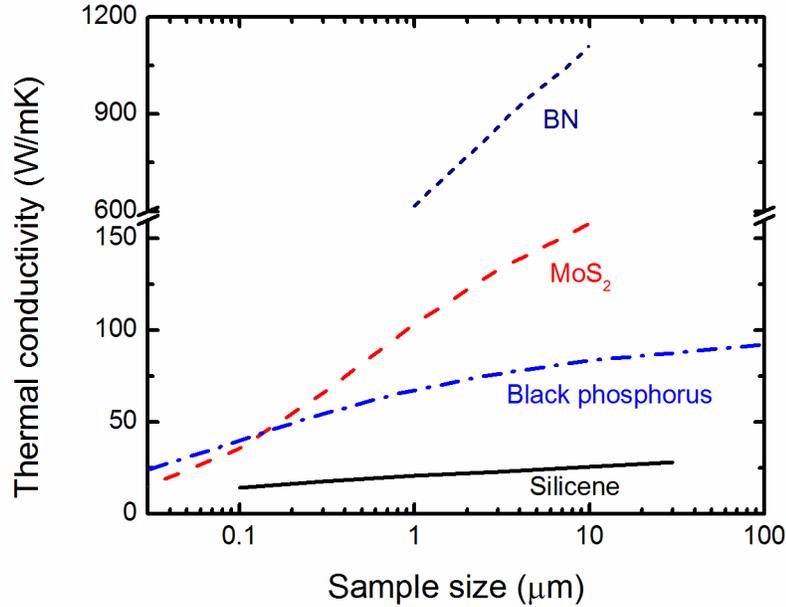

FIG. 14: Length-dependent thermal conductivity of some 2D materials from PBTE calculations, as reported by Gu and Yang for silicene [54] and MoS$_2$ [55], Lindsay and Broido [111] for h-BN and Zhu *et al*. [148] for black phosphorus along the zigzag direction.

## 5.2  Thermal conductivity of an infinitely large 2D crystal

Let's now discuss whether the thermal conductivity of infinitely large 2D sheets are finite? While the thermal conductivity of 3D bulk materials is finite and that of 1D system is generally believed to be divergent with the length [156], the question of whether the thermal conductivity for an infinitely-large 2D material is finite or unbounded is still controversial. Since the size of samples that experimental measurements can handle is always limited, theoretical analysis is expected to provide some insights to this problem.

Here, we consider phonon transport in infinitely large 2D sheets under the SMRTA. Considering that the SMRTA which uses phonon lifetimes as phonon relaxation time gives the lower bound value of the thermal conductivity [157], we would be able to conclude that the thermal conductivity from a more accurate iterative solution of PBTE is unbounded if the thermal conductivity is found to be unbounded under the SMRTA. Whether the thermal conductivity of a 2D crystal is finite depends on whether the integral of Eq. (16) is finite. As the integrand of Eq.

(16) is finite except at the Γ point (**q** = 0), to obtain a finite thermal conductivity requires that the long-wavelength phonons on each phonon branch satisfy the condition that $v_{\mathbf{q}s}^2 \omega_{\mathbf{q}s}^2 n_{\mathbf{q}s}^0 (n_{\mathbf{q}s}^0 + 1) \tau_{\mathbf{q}s}^{ph} \propto q^n$ with $n > -2$. As the group velocities of the long-wavelength optical phonons are zero, the long-wavelength acoustic phonons determine whether the thermal conductivity of the material is finite or not. While the in-plane acoustic phonons follow the $\omega_{TA,LA} \propto q$ relation near the Γ point, the flexural acoustic phonon branch become $\omega_{ZA} \propto q^2$ or $\omega_{ZA} \propto q$, depending on whether the 2D crystal is an unstrained one-atom-thick sheet or not. A finite thermal conductivity requires that the relaxation time of long-wavelength acoustic phonon be $\tau_{\mathbf{q}s}^{ph} \propto q^n$ with $n > -2$ if the dispersion is linear, or $n > -4$ if the dispersion is quadratic.

To analyze the scaling relation between phonon relaxation time and the wavevector, we classify the 2D crystals to three categories, unstrained one-atom-thick crystals, strained one-atom-thick crystals and the non-one-atom-thick crystals. The scattering mechanisms for long-wavelength acoustic phonons are quite different in each of these three cases.

The flexural acoustic phonon branch is generally believed to be quadratic for unstrained one-atom-thick 2D materials, such as graphene and h-BN. Bonini *et al*. [92] analytically derived the scattering rates for both in-plane acoustic phonons and flexural acoustic phonons. They found that the in-plane acoustic phonons are dominantly scattered through the decay processes into two flexural acoustic phonons due to the quadratic shape of the ZA branch, and the corresponding scattering rate is a constant value. ZA phonons also scatter with in-plane acoustic phonons in a reverse manner: they annihilate with an in-plane acoustic phonon and generate another in-plane acoustic phonon. The scattering rate of such events scales as $q^{-2}$. Therefore, the scattering rates of both in-plane acoustic phonons and flexural phonons satisfy the condition to ensure a finite thermal conductivity. The finite thermal conductivity of unstrained graphene is observed by Pereire and Donadio in their EMD simulations [103], where they found that the thermal conductivity is converged as the size of simulation domain increases. However, Lindsay [158] pointed out that the thermal conductivity becomes unbounded through the iterative solution of PBTE, which reflects the process of the redistribution of phonon modes driven by the heat flux. The underlying mechanisms are still not well understood. Since the main scattering mechanism for long-wavelength ZA phonon is the normal process, ZA+ZA<->LA/TA, the unbounded thermal transport could be related to the fact that a large portion of in-plane phonons are converted to ZA phonons.

When a tensile strain, even an infinitesimal one, is applied to the pristine 2D materials with a quadratic dispersion, the dispersion of the ZA branch becomes linearized [92]. In this case, the ZA phonons cannot be as efficiently scattered as the unstrained 2D sample and the scattering rate of ZA phonons scales as $q^{-3}$ [92]. Thus, the thermal conductivity of a strained sample becomes divergent as the length of the sample increases.

For non-one-atom-thick 2D materials, where the dispersion of ZA phonons is linear, no the symmetry selection rule exists for ZA phonons. Therefore, the scattering of ZA phonons is much more frequent than strained one-atom-thick 2D materials as the scattering rate of ZA phonons scales as $q^{-1}$, and the origin for the unbounded thermal conductivity does not come from the ZA phonons as the strained one-atom-thick 2D materials. Gu and Yang [54] examined the length-dependent phonon transport in silicene, a non-one-atom-thick 2D material with hexagonal lattice, whose lowest acoustic phonon dispersion is linear around the Γ point,. They found that the long-wavelength LA phonons cannot be effectively scattered to make the scattering rate satisfy the condition that $q^n, n > -2$. They further decomposed the scattering of acoustic phonons into different channels. The decomposed scattering rates of phonon modes along high-symmetry direction (from the Γ point to the M point) are shown in Fig. 15. The scattering of a long-wavelength acoustic mode can be classified as four conditions: the decay process to two lower-frequency modes on the same branch, the decay process to two modes on the lower acoustic branches, the annihilation process with modes on the same branches and the annihilation process with two modes on the different branches. From analytical derivation, only the annihilation process with two modes on the same branches could make the scattering rate follow $q^n, n > -2$. For a long-wavelength LA phonon $\mathbf{q}s$, its frequency can be expressed by $\omega_{\mathbf{q}s} = v_{LA}q$, with the sound velocity of LA branch $v_{LA}$. The frequencies of the two phonons, $\mathbf{q}'s'$ and $(\mathbf{q}'+\mathbf{q})s'$ on the same branch that can scatter with $\mathbf{q}s$ have to satisfy the condition, $\omega_{(\mathbf{q}'+\mathbf{q})s'} - \omega_{\mathbf{q}'s'} = v_{LA}q$. As $q$ approach zero, the condition becomes $\mathbf{v}_{\mathbf{q}'s'} \cdot \mathbf{q} = v_{LA}q$. However, since the sound velocity of LA branch is the larger than the group velocity of other phonon modes, long-wavelength LA phonons cannot be scattered by two phonons on the same branch. Thus, the thermal conductivity of silicene is divergent due to the LA branch.

    We note that the above analysis only takes into account the three-phonon scattering processes. At high temperature higher-order phonon-phonon scattering might be important. There yet exists a study on whether the higher-order scattering mechanism could change the above conclusion. Therefore, it is desirable to have more theoretical studies on this topic, including MD simulations with longer sample length and PBTE calculations with higher-order scatterings.

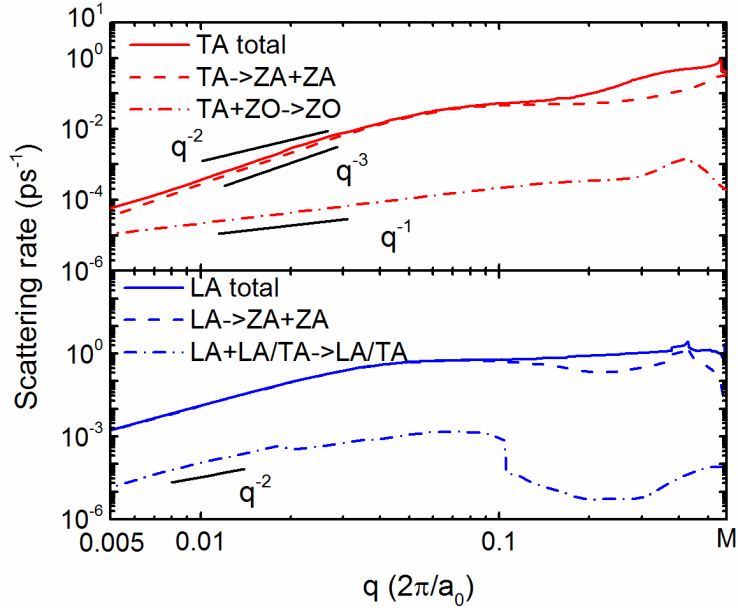

Figure 15. Scattering rates of in-plane LA and TA acoustic phonons in silicene along Γ-M direction at 300 K from the first-principles calculations [54].

## 5.3 Thermal transport in nanoribbons

Nanoribbons exhibit very different electronic properties than the large 2D crystals. For example, the bandgap of the graphene can be opened when it is patterned into nanoribbons. Novel electronic devices using nanoribbons have been fabricated [159, 160]. The thermal conductivity of nanoribbons of 2D crystals has also been intensively studied. Low thermal conductivity of nanoribbons with rough edge could potentially find their applications in thermoeletrics [161] while asymmetric nanoribbons exhibit thermal rectification effect like a diode to control the heat flow [162, 163].

Similar to the nanostructures of bulk crystals, such as nanowires and thin films, the boundaries of nanoribbons can scatter phonons and thus reducing the thermal conductivity. The thermal conductivity of nanoribbon usually decreases with the decrease of the nanoribbon width from MD [101, 102] and PBTE [98, 100]，which is mainly attributed to stronger phonon-boundary scattering in narrower ribbons.

Bae *et al*. [153] measured the thermal conductivity of $SiO_2$-supported graphene nanoribbons with different length and width using the micro-bridge method. It was found that at the room temperature the thermal conductivity of the graphene nanoribbon samples with a length $L$=260 nm increases rapidly with the increasing of the width when the width is below 100 nm, where edge boundary scattering dominates the phonon transport, and gradually converges to around 300 W/mK They proposed a scaling relation between the thermal conductivity $\kappa$ and the width of nanoribbon $W$, $\kappa \propto W^{1.8\pm0.3}$, to characterize the thermal transport in the edge-limited regime.

In addition to the width of nanoribbons, several other factors can also affect the thermal conductivity, including the edge chirality, roughness and passivation [98, 101, 102, 164, 165]. Zigzag nanoribbon has a larger thermal conductivity than that of armchair nanoribbons with the same width. Some studies attributed it to the different phonon spectra of the zigzag and armchair nanoribbons [166], while others speculated that phonon-boundary scattering is strongly dependent on the type of edge chirality [162]. By modeling nanoribbons as 1D crystal in AGF calculations [166, 167], phonon dispersion of nanoribbons was computed. It was found that the low-frequency phonon branches in zigzag nanoribbons are more dispersive than those in armchair nanoribbons, leading to larger group velocities and thus larger thermal conductivity. In the phonon-boundary scattering picturing [102, 162], the boundary scattering is expected to be weaker in zigzag nanoribbons since the edge roughness of the zigzag nanoribbons is smaller than the armchair ones. Wei *et al*. [168] used wave-packet simulations to study the strength of boundary scattering events and found that the zigzag edges provide little resistance to heat flow since phonons are always specularly scattered at the zigzag edges, but wave packet splitting and mode conversion occur at the armchair edges. Apart from these two explanations, Wang *et al*. [169] argued that the difference of the roughness in the two kinds of edge roughness is not enough to explain the different degrees of suppression on the thermal conductivity. Stronger localization of phonons happens closer to the armchair edge region than that of zigzag edge region.

Aside from the zigzag and armchair edges, the thermal conductivity of nanoribbons could be tuned by introducing larger scale edge roughness [101]. For example, a rough nanoribbon with a width of 2 nm has only 1/6 the conductivity of its smooth counterpart. Aksamija and Knezevic [98] used the ratio of root-mean-square height of edge variations to phonon wavelength to estimate the specularity parameter and calculate the thermal conductivity of nanoribbons with different roughness.

MD simulations have also been carried out to study the thermal conductivity of $MoS_2$ nanoribbons. Unlike graphene, melting was observed at the edge of $MoS_2$ nanoribbons by Jiang *et al*. [50]. Liu *et al*. found that the thermal conductivity is almost independent on the width of nanoribbons [142]. This might be due to the strong anharmonicity in the empirical potentials employed in their MD simulations.

## 5.4 Layer-dependent thermal transport

It is also of fundamental and practical importance to study layer-dependent thermal transport in 2D materials, stimulated by the layer-dependent electronic and optical properties in 2D materials. One might expect the thermal conductivity of thicker few-layer 2D material is higher than that of the thinner ones, just as the size-dependent thermal transport in nanoribbons, due to the stronger phonon-boundary scatterings. However, due to the weak *van der* Waals interaction in the layered 2D crystals, the strength of boundary scattering, which is approximated as $\Gamma_{\mathbf{q}s}^{B} \propto |v_{\mathbf{q}s}^{z}|/t$ with the thickness *t*, is expected to be much weaker than conventional thin films, such as silicon thin films. In addition, when the thickness is reduced to the order of 1 nm, the phonons are confined in the 2D plane, the propagation perpendicular to the basal-plane direction is totally forbidden, thus the boundary scattering do not exist. Therefore, thickness-dependent thermal conductivity in 2D crystals could be quite distinct from that of conventional thin films.

Figure 16 summarizes the study on the thermal conductivity of graphite, graphene and the multi-layer graphene using different methods. On the experimental side, the thermal conductivity of graphite is about 2000 W/mK at room temperature, while that of graphene ranges from 2500-5000 W/mK. The systematic studies using both MD and PBTE also confirmed that single layer graphene has a much higher thermal conductivity. As discussed in Sec. 4.1, this is because the symmetry selection rule is broken in single layer graphene [62], which leads to weaker scattering for the flexural phonon modes in graphene. Experiments [27] and numerical simulations [90, 170, 171] further show that thermal conductivity gradually decreases when the layer number is increased. For multi-layer graphene, there are additional ZA-like low-frequency optical phonon branches generated compared with the single-layer graphene. It was found that the reduction of thermal conductivity is mainly due to less heat transported by ZA and ZA-like phonon modes [170]. Compared to the ZA modes in single-layer graphene, these ZA-like phonons in multilayer graphene have lower group velocities and much larger phonon scattering rates due to more scattering channels available. Similar monotonic reduction of thermal conductivity with respect to number of layers was also observed for h-BN from PBTE-based calculations [112].

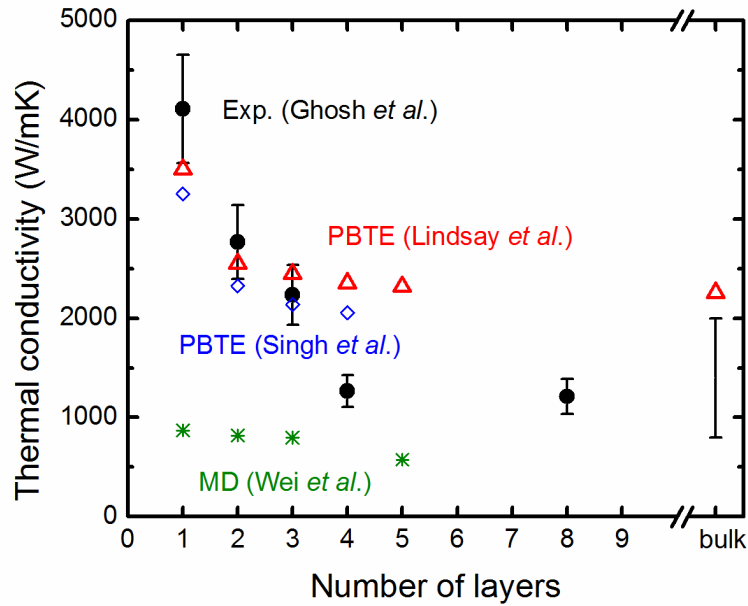

FIG. 16: The thermal conductivity of graphene/graphite reported by Ghosh *et al.* [27] from optothermal Raman measurements, Lindsay *et al.* [170] from PBTE calculations, Singh *et al.* [90] from PBTE calculations, Wei *et al.* [171] from MD calculations.

Recent experiments show a quite different trend for the layer-dependent thermal conductivity of $MoS_2$. Thermal conductivity increases with the layer number, as shown in Fig. 17. The measured value of the thermal conductivity of single-layer $MoS_2$ is 34.5 W/mK [137], then increases to 46 W/mK, 50 W/mK, 52 W/mK for 4-layer, 7-layer [138] and 11-layer $MoS_2$ [29]. Liu *et al.* [140] reported the thermal conductivity for bulk $MoS_2$ is around 110 W/mK. As discussed by Jo [138], since the samples are different and the experiment conditions are different, the direct comparison among these data might be questionable. It is highly desirable to have the thermal conductivity of the samples with different layer numbers measured using the same experiment setup. In addition, polymer residues and absorbed gas might influence the thermal conductivity results. Since the fewer-layer samples are more sensitive to these impurities than the thicker ones, the layer-dependence of the intrinsic thermal conductivity might be quite different from that of the contaminated ones.

The layer-dependence of the thermal conductivity of h-BN seems to follow the same trend as $MoS_2$ [39]. The room-temperature thermal conductivity of suspended h-BN can approach the basal-plane values of bulk h-BN crystals when the thickness is increased to more than 10 atomic layers in spite of the presence of polymer residue on the sample surface. As the sample thickness decreases, the thermal conductivity decreases because of increasing phonon scattering by polymer residues, which is more pronounced at low temperatures or for low-frequency phonons.

Yan *et al.* [172] measured the thermal conductivity of another 2D transition metal dichalcogenide $TaSe_2$ with 1T structure using the optothermal Raman measurement. They observed an opposite trend for $TaSe_2$ with 1T structure than $MoS_2$ with 2H structure. The thermal conductivity of these $TaSe_2$ films at room temperature decreases from its bulk value of 16 W/mK to 9 W/mK in 45-nm-thick films. Such dependence of thermal conductivity on the film thickness suggests that phonon scattering from the film boundaries could be substantial despite the sharp interfaces of the mechanically cleaved samples.

Qiu and Ruan [147] studied the thermal conductivity of $Bi_2Te_3$ with different layers using EMD simulations with the empirical potential they developed. They found a non-monotonic dependence of the thermal conductivity on layer thickness. The single-layer $Bi_2Te_3$ has the highest thermal conductivity, which is then reduced to the minimum value for the three-layer $Bi_2Te_3$, and then increases and converges back to the bulk value. The authors attributed this behavior to the competition between interfacial scattering and the Umklapp phonon-phonon scattering.

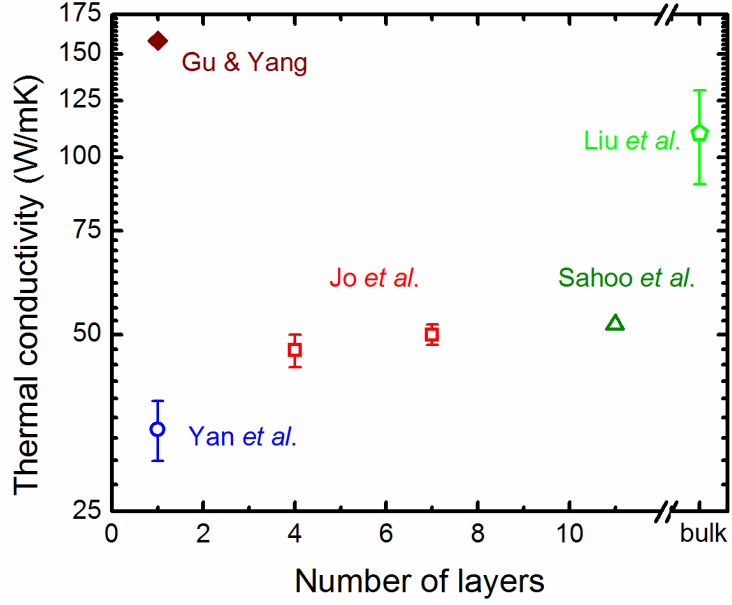

FIG. 17: Measured thermal conductivity of single-layer, few-layer and bulk MoS$_2$: Yan *et al*. [137] for single-layer sample, Jo *et al*. [138] for 4-layer and 7-layer samples, Sahoo *et al*. [29] for 11-layer sample and Liu *et al*. [140] for bulk sample. First-principles prediction for a 10-μm-long single layer MoS$_2$ by Gu and Yang [55] is also plotted for reference.

In short, the layer number-dependent thermal transport in 2D materials is much more complicated than expected from the thickness-dependence of thin films. Different trends of thermal conductivity have been observed as the layer number increases, which is worth further investigation.

# 6 Effects of defects and strain

## 6.1 Isotope and defects

Imperfections, such as point defects, vacancies and grain boundaries, inevitably occur in 2D materials, especially the CVD-grown ones. These imperfections usually lead to phonon scattering and thus reducing the thermal conductivity.

Theoretically, the phonon scattering rate due to point defects can be derived from the Fermi's golden rule [173] or calculated using the Green's function approach [80, 81]. For 2D crystals, the phonon scattering rate can be roughly estimated by [60, 174, 175]

$$1/\tau_{\mathbf{q}s}^{D} \propto \left(\omega^{3}/v^{2}\right) \sum_{i} f_{i} \left[ \left(1 - m_{i}/\bar{m}\right)^{2} + \varepsilon \left(\gamma \left(1 - r_{a,i}/\bar{r}_{a}\right)\right)^{2} \right] \quad (21)$$

where $M_i$, $r_{a,i}$ and $f_i$ are the mass, radius and fraction concentration of type $i$ atom, $\bar{M}$ and $\bar{r}_a$ is the averaged atom mass and radius, $\gamma$ is the Gruneisen parameter, $v$ is the phonon group velocity and $\varepsilon$ is a phenomenological parameter characterizing the relative importance of defect scattering due to mass difference and lattice distortion. The expression indicates that the high-frequency/short-wavelength phonon modes are more likely to be scattered by defects. Chen *et al*. [28] synthesized the graphene sheets with different compositions of $^{12}$C and $^{13}$C, and measured their thermal conductivities. A small concentration (1.1%) of isotopes leads to a 30% reduction of thermal conductivity. For a 50%/50% concentration, the thermal conductivity is reduced to 50% of the graphene with 100% $^{12}$C. Chen *et al*.'s observation is consistent with their MD simulations. The isotope effects were also explored in other 2D materials, such as h-BN and silicene. Lindsay *et al*. [111] found that the thermal conductivity of a 10-μm-long isotopically-pure h-BN sheet can be 37% higher than the same size h-BN sheet made up of naturally-occurring B and N isotopes.

Apart from randomly distributed isotopes, Mingo *et al*. [106] proposed to use isotope clusters to scatter phonons and to reduce the thermal conductivity. As the scattering is sensitive to the dimension of imperfections, the isotope clusters serve the same idea as the nanoparticles in bulk materials [176] to scatter low-frequency phonons and thus further reducing the thermal conductivity. In Mingo *et al*.'s AGF calculations, isotope clustering is found to result in a remarkable increase of the scattering cross-section and thus a significant reduction on the thermal conductivity.

Equation (21) clearly indicates that introducing different kind of atoms could more severely scatter phonons, since it not only introduce the mass difference, but also the lattice distortion. Due to its 2D nature, the distortion in 2D materials could be more distinct from that in 3D bulk material. A study was carried out to investigate how phonons transmit across the Si defect in graphene. Due to a much longer Si-C bond than a C-C bond, the Si defect leads to ripples [177]. AGF study shows a smaller phonon transmission across this defect and thus a largely reduced thermal conductivity is expected.

Aside from isotopes and substitutions, vacancy defect where one or some atoms are missing, also scatter phonons [104, 178]. Compared to isotopes, vacancies can more effectively suppress the thermal conductivity. For example, MD simulations show that the thermal conductivity of graphene with 1% vacancies is reduced to 10-20% of the unmodified graphene [178], but only a 30% reduction is found when 1% $^{12}$C atoms are replaced by $^{13}$C atoms [28]. From the classical theory, if a vacancy is modeled as removing one atom from the crystal and all the linkages associated with this atom, the effect of vacancy on phonon scattering rate can be estimated by [179]

$$1/\tau_{\mathbf{q}s}^{\text{v}} \propto \frac{\omega^4}{4\pi v^3} \tag{22}$$

where $G_v$ is the number of atoms per defects. Recently, Xie *et al*. [180] pointed out that the classical model overlooks the different properties of bonds connecting under-coordinated atoms, which are shorter and stronger than ordinary bonds. They proposed a modified model to take the bond-order into account. This model can then be incorporated with PBTE formalism to predict the thermal conductivity of 2D crystals with defects.

Grain boundaries occur in polycrystalline structures, where crystals with different crystal orientations are connected together. The grain boundary can be regarded as an array of vacancies.

Depending on the tilt angle between two pieces of 2D sheets, the density and type of vacancies are different [181] and the overall thermal transport properties are altered [182-185]. **Bagri** *et al*. [182] calculated the thermal boundary conductance across a grain boundary, ranging from 15 to 45 GW/m$^2$K using a modified Tersoff potential [74]. They pointed out that the effect of the tilt angle is important that the response of boundary conductance to the tilt angle is comparable to the intrinsic conductance of a grain with a size smaller than 100 nm. Furthermore, AGF calculations [184] provide more detailed information on phonon transmission across the grain boundaries, but the calculated value of boundary conductance is several times smaller than that from MD simulations. This is because the anharmonicity is ignored in AGF, which is expected to facilitate the phonon transport across the interface by converting some phonon modes to others.

Large scale MD simulations were also conducted to calculate the effective thermal conductivity of polycrystalline graphene sheets with an averaged grain size of 1-5 nm. The thermal conductivity was found to be one order of magnitude smaller than that of pristine graphene [107]. In a similar study, the thermal conductivity of polycrystalline graphene was shown to decrease exponentially with increasing grain boundary energy [108]. In another work, the thermal conductivity in large-area (10 μm × 10 μm) polycrystalline graphene was studied [99], where the thermal conductivity values of small grains with a rough grain boundary were computed by PBTE approach and then used to generate temperature distribution over the large-scale sample based on the heat flux continuity for each grain.

## 6.2 Strain

Strain has been used to tune electronic properties of many materials including 2D crystals [186, 187]. It is of great interest to integrate strained 2D materials into nanoscale electronics and photonics. In heterostructures, where two or more kinds of 2D crystals are vertically stacked or laterally connected together, strain always exists due to the lattice mismatch [188]. Therefore, it is desirable to understand the role of strain on thermal transport in 2D materials [189]. Strain alters both phonon dispersion [190] and the rate of phonon scattering. Strain could also lead to defects and instability in the crystal. All these could in turn change the thermal conductivity. There have been a few studies on thermal transport in 2D materials under strain.

Li *et al*. [76] performed EMD simulations for graphene and showed that the thermal conductivity has a non-monotonic dependence on strain and the highest thermal conductivity occurs at the zero strain condition. The authors attributed the decrease of the thermal conductivity with respect to the compressive strain to the rippling of graphene induced by the compression since the deformation of graphene could lead to additional scattering to phonons in addition to phonon-phonon scatterings. When applying tensile strain, the phonon dispersion shifts downward, leading to lower group velocity and the mode-wise heat capacity. The nono-monotonic trend is also confirmed by Guo *et al*. [102] and Wei *et al*. [191] in graphene nanoribbons. However, the role of strain on the thermal conductivity of graphene is not conclusive when graphene is stretched. EMD simulations by Pereira and Donadio [103] on uniaxially-strained graphene showed that the thermal conductivity of the uniaxially-stretched graphene with 1% strain along the strain direction is close to the unstrained graphene, but it becomes divergent when the strain is increased to 2% or larger.

Bonini *et al*. [92] solved the PBTE under SMRTA, and found that the thermal conductivity of an infinite graphene sheet changes from a finite value to infinity when a small strain is applied. The applied strain changes the dispersion of the ZA branches from quadratic to linear relation and the scattering rates of long-wavelength ZA phonon modes scales as $q^{-3}$, which renders an infinite thermal conductivity due to the ZA phonon modes, as discussed in Sec. 5.1. The findings from the PBTE calculations with RTA seem to indicate that tensile strain increases the thermal conductivity. However, when solving PBTE with the more accurate iterative approach, the strain effects on the thermal conductivity become more complicated. Lindsay *et al*. [66] found that a 1% tensile strain results in a slight reduction on the thermal conductivity of a 10-μm-long graphene, while in Fugallo *et al*. [91] showed that the thermal conductivity could be either increased or reduced depending on the sample size and the amount of strain applied. In particular, the reduced thermal conductivity is found when a small strain is applied on a large size sample, but the thermal conductivity always increases with the strain when the strain is larger than 2%. The different results are obtained from the SMRTA and the iterative solutions for the large size samples, when the boundary scattering is weak comparing to the intrinsic scattering mechanisms. This difference is likely due to the fact that normal scattering was not correctly taken into account in the RTA. It is thus desirable to have more systematic analysis to understand the effect of strain on phonon transport in graphene, in particular on the role of normal scattering and long-wavelength acoustic phonons.

The effects of tensile strain on the thermal conductivity of 2D crystal could be quite diverse depending on their lattice structures. The MD simulations with the MEAM potential showed that the thermal conductivity of silicene increases first with the tensile strain and then decreases [126]. The maximum thermal conductivity occurs at a 4% tensile strain. The authors attributed this behavior to the response of in-plane stiffness to the strain. They found that the phonon frequency becomes larger with the increase of strain, which is quite different from the conventional bulk materials and graphene. The MD simulations with a Stillinger-Weber potential by Jiang *et al*. showed that the thermal conductivity of $MoS_2$ decreases with the tensile strain [50], which is similar to bulk materials, where the phonon dispersion downshifts and thus the group velocity and heat capacity decreases.

Comparing to graphene, the studies of other 2D materials under compressive strain are very limited, likely because it is difficult for PBTE formalism to deal with the 2D sheets with rippling and there are limited potentials available for MD simulations on other 2D materials.

# 7   Device geometry

The heat dissipation capability of 2D materials in practical applications could be quite different from the freestanding ones when they are integrated into device geometry. Here we summarize two common situations: 1) 2D materials are placed on a substrate, where heat is dissipated through both in-plane thermal conduction and leakage into the substrate through the contact. 2) heterostructured 2D materials.

## 7.1 Effect of Substrates

When 2D materials are integrated for device applications, thermal properties could be significantly altered due to its contact with substrates. Seol *et al*. [34] measured the thermal conductivity of supported graphene, which is only around 600 W/mK in comparison with ~3000 W/mK for suspended graphene. PBTE was used to model the heat transport by modeling phonon-substrate scattering as point contacts. Ong and Pop [192] found that the thermal conductivity of supported graphene coupling to $SiO_2$ substrate is reduced by an order of magnitude, due to the damping of the flexural acoustic (ZA) phonons. However, increasing the strength of the graphene-substrate interaction enhances the heat flow and the effective thermal conductivity along the supported graphene. The enhancement is due to the coupling between the ZA modes in graphene and the substrate Rayleigh waves, which increases the group velocity of the hybridized modes. Qiu and Ruan [193] conducted similar simulations using EMD, and the spectral energy density analysis showed that the relaxation time of acoustic phonons and optical phonons were reduced to 1/10 and 1/2 of the values for suspended graphene. The thermal conductivity of graphene nanoribbon placed on SiC substrate was also investigated using MD simulations [194] where the interfacial interaction was modeled as both the covalent bonding and *van der* Waals bonding. The single-layer graphene nanoribbon has a much lower thermal conductivity when it is covalently bonded with the substrate than that of the weakly bonded *van der* Waals interaction. However, if a bi-layer graphene were bonded to substrate, the second layer of graphene nanoribbon maintains the high thermal conductivity. Due to the suppression of the ZA modes which conduct most of the heat in graphene, the length-dependence of the thermal conductivity of graphene becomes weaker [195].

Since the thermal conductivity of the 2D materials is suppressed when it contacts with substrate, it is a desirable to identify the thickness (layer number) beyond which the thermal conductivity can recover to a high value as the freestanding sample. Jiang *et al*. [196] experimentally measured the thermal conductivity of graphene and ultrathin graphite (thickness from 1 to 20 layers) encased within silicon dioxide using a heat spreader method. In contrast to the negative layer number-dependence of the thermal conductivity of suspended few-layer graphene, the thermal conductivity of the encased graphene increases with the number of layers. To maintain the high thermal conductivity (>1000 W/mK), the thickness of encased graphene should be larger than 10 nm. Sadeghi *et al*. [197] conducted more experiments on the thermal conductivity of few-layer graphene supported on $SiO_2$ substrate. The full recovery of the thermal conductivity of a few-layer graphene on $SiO_2$ substrate to the thermal conductivity of graphite requires the thickness of few-layer graphene to be beyond 34 layers. This is probably due to long phonon mean free path along the cross-plane direction, which was verified by recent MD simulations [198] and experimental measurements [199], as well as due to the partially diffusive interface-phonon scattering, whose specularity parameter of graphite/$SiO_2$ interface ranges between 0 and 0.36 according to Sadeghi *et al*. [197].

Heat could transport across the interface of 2D material and substrate and subsequently dissipate to the substrate. The interfacial thermal conductivity between graphene and a wide range of substrates including graphite have been measured using the pump-probe thermoreflectance method, micro-bridge method and optothermal Raman techniques, as summarized in Ref. [18]. The typical thermal conductance between graphene and substrate materials ranges from 20 to 100 MW/m$^2$K at room temperature.

There exist much fewer measurements on the interfacial thermal conductance of other 2D materials. Figure 18 shows the available data on h-BN and MoS$_2$. In general, the interfacial thermal conductance is smaller than that of graphene. Jo *et al*. [39] extracted both the thermal conductivity of h-BN and the interfacial thermal conductance from their micro-bridge experiment. As the sample was transferred by PMMA, the residues were thought to be responsible for the low thermal conductance. Liu *et al*. [140] showed that the interfacial thermal conductance between MoS$_2$ and some metals are smaller than the interface of Al film and highly ordered pyrolytic graphite. Taube *et al*. [139] extracted the interfacial thermal conductance of single-layer MoS$_2$ and SiO$_2$ support from optothermal Raman measurement and found that the value is comparable to h-BN on SiN$_x$ substrate.

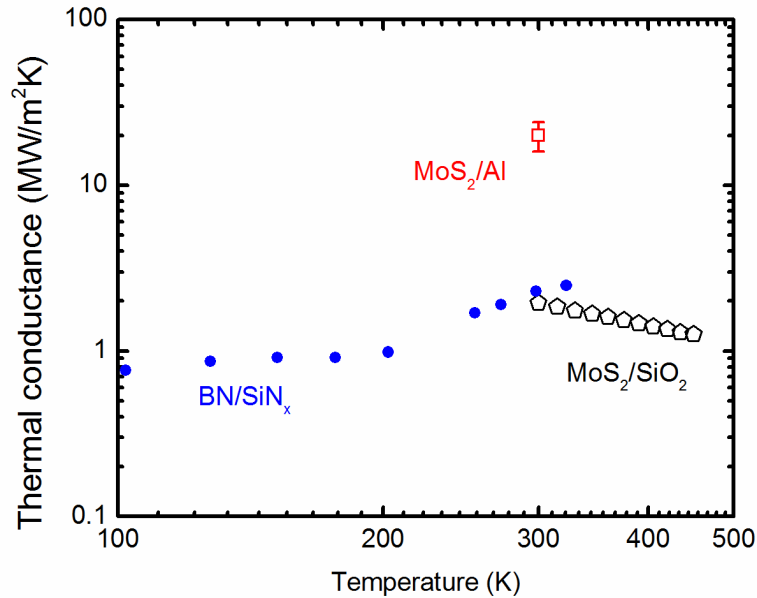

FIG. 18: Temperature-dependent interfacial thermal conductance for h-BN/SiN$_x$ interface reported by Jo *et al*. [39], MoS$_2$/Al interface reported by Liu *et al*. [140], MoS$_2$/SiO$_2$ interface reported by Taube *et al*. [139].

## 7.2 Phonon transport in heterostructures

Many efforts are devoted to synthesize heterostructures for electronics and optoelectronics applications based on a diverse set of 2D materials. Both in-plane and vertical heterostructures of TMDs have been synthesized [200, 201]. The thermal transport in such heterostructures has yet been systematically studied, though there are a few studies on graphene/h-BN and graphene/silicene related structures through MD simulations [202-205] and AGF approach [206].

MD simulations showed that the thermal conductivity of in-plane graphene/h-BN superlattice along the direction of graphene/h-BN interface is sensitive to the type of the interface. If the graphene and h-BN stripes are connected by zigzag edge, the thermal conductivity is larger than that connected by the armchair edge [110]. The effect is pronounced when the width of the stripe is small, but both of them are found to converge to the average thermal conductivity of graphene and h-BN. When the heat flux is perpendicular to the interface, the thermal conductivity of in-plane superlattice of 2D crystals decreases with the decease of superlattice period first, then increases when reducing further the period [202]. A critical period of 5 nm was found to lead to the minimum thermal conductivity. The dependence of the thermal conductivity of 3D superlattice on the period has been discussed [207, 208]. The first reduction is mainly due to increased interface scattering (more interfaces), then recovery is attributed to coherent effect, such as the opening of phonon band gaps and reduction of group velocity.

Thermal transport in graphene with embedded hexagonal h-BN quantum dots has been studied using both a real-space Kubo approach [209] and EMD simulations [110]. Both studies showed that the thermal conductivity of graphene with quantum-dot nanostructures is reduced compared with the pristine graphene. The reduction of the thermal conductivity is less pronounced with larger quantum dots. Minimal thermal conductivity was found at 50% concentration for different radii 0.495, 1.238, and 2.476 nm of h-BN quantum dots.

The thermal conductivity of vertical graphene/silicene heterostructure has been studied, where silicon thin film or single layer silicene is sandwiched between two graphene monolayers [204]. The thermal conductivity of the heterostructure decreases as the silicon film thickness increases, where interfacial interaction between graphene and silicon/silicene are modeled by either strong covalent bonding or weak *van der* Waals interactions. For the *van der* Waals interaction, the thermal transport is mainly contributed by the flexural modes for thin Si film, while the longitudinal modes dominate for thick Si film. For covalent bonding, phonon transport is simply governed by the longitudinal modes. The contribution from the longitudinal modes increases as Si film thickness increases. The thermal conductivity of vertical graphene/$MoS_2$ heterostructure is reported to be 1037 W/mK from a MD study [210].

The thermal conductance across interfaces between two different 2D materials in either vertical or in-plane heterostructure has also been investigated. Chen *et al*. [211] measured the interfacial thermal conductance between graphene and h-BN, which are vertically stacked, using optothermal Raman technique and found that the interfacial thermal conductance is about 7.4 MW/$m^2$K at room temperature. They attributed the low interfacial thermal conductivity for h-BN/graphene to the contaminations of the interface by liquid or organic residues. In addition, they expected that the interfacial thermal conductance can be enhanced if lattice-matched graphene is grown on h-BN. MD simulations with a transient heating scheme, which mimics the laser heating in the Raman measurement, were also performed and an interfacial thermal conductance value of 3.4 MW/$m^2$K were found, which is consistent with the Raman experiment [212]. The simulations further showed that the interfacial thermal conductance can be enhanced by raising the temperature, increasing bonding strength or introducing graphene hydrogenation. A similar transient MD study showed that the interfacial thermal conductance value for graphene/$MoS_2$ interface is 5.8 MW/m$^2$K [210]. NEMD simulations were performed for the interfacial thermal conductance of an in-plane interface between graphene and silicene where a value of 250 W/m$^2$K was obtained [213].

# 8 Chemical tuning

Thermal conductivity of 2D materials can also been tuned through surface functionalization and intercalation.

## 8.1 Surface functionalization

Some 2D materials can be made by chemical functionalization of existing materials. Many graphene derivatives have been synthesized through hydrogenation, oxidization and fluorination. Table VI summarizes thermal transport properties of 2D graphene derivatives studied using atomistic simulations.

Compared with pristine graphene, the calculated thermal conductivities of hydrogenated graphene (graphane) and fluorinated graphene (flurographene) with a 100% coverage of the H and F atoms are reduced to 40%-50% [214-216] and 35% [217] from MD simulations and ~50% and ~7% from a first-principles-based PBTE study [91], respectively. The phonon lifetime time of graphane extracted from MD simulations was found to be decreased by an order of magnitude compared with pristine graphene [215]. The larger scattering rate is regarded to come from the coupling between the in-plane and out-of-plane phonon modes [218].

Unlike the conventional thin film materials, the thermal transport in the functionalized graphene can be further tuned through the coverage of the functionalized molecules. When the coverage increases from 0% to 100%, the thermal conductivity generally exhibits a U-shape curve [214, 217, 218], similar to the dependence of alloy thermal conductivity on composition. Mu *et al*. [154] showed that graphene oxide with a coverage of 20% has a thermal conductivity lower than the calculated thermal conductivity of graphene according to minimum thermal conductivity theory [219].

## 8.2 Intercalation

There has been a rich chemistry in intercalating 2D materials to form composites or superlattice of layered 2D materials [220-222]. Intercalation compounds are formed by the insertion of atomic or molecular layers of a different chemical species called the intercalant between layers in a layered host material, such as graphite, transition metal dichalcogenides, and other *van der* Waals crystals. When the intercalants are inserted into the gap of layered materials, changes can be transferred to the layers of the host materials[222], leading to the change of the Fermi level and density of states close to the Fermi level [221]

Besides the electrical properties, the thermal conductivity also changes significantly due to intercalation. The study of thermal conductivity of intercalated compounds can be traced back to 1980s [223-225]. The thermal conductivities of graphite intercalation compounds with different kinds of intercalants, such as $FeCl_3$ and $SbCl_5$, were measured. It was shown that intercalation generally leads to an increase of carrier density and thus the electronic thermal conductivity, and a reduction of the lattice contribution. As low temperature (<4 K), electronic part dominates the total thermal conductivity, while at room temperature the lattice part dominates. As a results, the thermal conductivity is enhanced at low temperature but reduced around room temperature [223].

To explain the lower lattice thermal conductivity around room temperature, Issi *et al*. [223] employed the PBTE formalism, which considered the scattering of in-plane phonon modes in graphene layers of the intercalation compound, including phonon-phonon scattering, grain boundary scattering and defects scattering. By fitting the experimental data with the analytical expressions of phonon relaxation times, they attributed the reduced thermal conductivity to the lattice distortion associated with large-scale defects due to intercalation process and the grain boundary scattering. However, the extracted feature size of the grains is about one order of magnitude smaller than the experimental observation. This indicates that there are additional scattering mechanisms that are not considered in their model.

Although many other intercalation compounds aside from graphite intercalation compounds have been synthesized, there are only limited thermal conductivity measurements.

$TiS_2$-based intercalation compounds have attracted considerable attention due to the large thermoelectric power of bulk $TiS_2$ crystals [211]. Wan *et al*. [226-229] intercalated different intercalants, such as SnS and BiS, into the $TiS_2$ gap to form superlattices and demostrated improved thermoelectric figure of merit. One example is the superlattice $(SnS)_{1.2}(TiS_2)_2$, where SnS layer is intercalated into $TiS_2$ layered crystal. The cross-plane thermal conductivity is shown to be lower than the predicted minimum thermal conductivity. The authors attributed the low thermal conductivity to phonon localization due to the orientation disorder of SnS layers [229].

Organic molecules can also be inserted into the *van der* Waals gaps. Recently, Wan et al synthesized the layered hybrid inorganic/organic material $TiS_2/[(Hexylammonium)_{0.08}(H2O)_{0.22}(DMSO)_{0.03}]$ for thermoelectric applications [230, 231]. Due to the low lattice thermal conductivity and the relatively unchanged power factor, the thermoelectric figure of merit ZT is as high as 0.42, making the hybrid material a promising n-type flexible thermoelectric material [230]. The measured thermal conductivity of the hybrid material is only one-sixth of the bulk value of $TiS_2$. More surprisingly, the lattice thermal conductivity was found to be only 1/30 of the bulk $TiS_2$. MD simulations were performed to calculate the thermal conductivity of the bulk and the hybrid materials, an eight-fold reduction on lattice thermal conductivity when the $TiS_2$ bulk is intercalated by organic components. This reduction is significantly higher than the simple volumetric averaging, indicating that the coupling between inorganic and organic parts plays an important role in phonon transport in the in-plane direction. They attributed the stronger scattering in the hybrid material to the strong ionic bonding between the inorganic $TiS_2$ layers and the dangling hexylammonium ions. In addition to the interlayer interaction, other scattering mechanisms such as the wavy structure of the layers and the local atomic disordering of the TiS$_2$ layers might also be responsible for the measured ultralow thermal conductivity.

## 9 Summary and Outlook

This article provides a summary on the exciting study of phonon transport and thermal conductivity of emerging 2D materials. The thermal conductivity of 2D materials spans about four orders of magnitude from a few W/mK to several thousand W/mK. The range of thermal conductivity can be tuned by controlling sample dimensions, applying strain, introducing defects, and surface functionalization and intercalating with guest molecules. The thermal transport properties in free-standing 2D materials could be quite different from those integrated into devices due to their interactions with substrates or other 2D materials in heterostructures.

Despite these many studies, there remain some questions and challenges. There are limited experimental techniques that could be used to characterize the ultra-thin 2D materials. Only a few research groups have the experimental platforms to characterize these materials. The thermal conductivity of many emerging 2D materials has yet been experimentally measured. The theoretical study on phonon transport in 2D materials has been limited by the availability of reliable experimental data. Furthermore, there has been very limited work on phonon transport in in device geometry and in complex heteostructures. Both modeling and experimentation techniques need to be developed to understand the full potential and benefits of emerging 2D materials.

## Acknowledgements

This work was supported by the NSF (Grant No. 0846561, 1512776) and DOD DARPA (Grant No. FA8650-15-1-7524). X.G. acknowledges the Teets Family Endowed Doctoral Fellowship.

## References

[1] K.S. Novoselov, A.K. Geim, S. Morozov, D. Jiang, Y. Zhang, S. Dubonos, I. Grigorieva, and A. Firsov, Electric field effect in atomically thin carbon films, *Science*, Vol. 306, pp. 666-669, 2004.
[2] K.S. Novoselov, A.K. Geim, S.V. Morozov, D. Jiang, M.I. Katsnelson, I.V. Grigorieva, S.V. Dubonos, and A.A. Firsov, Two-dimensional gas of massless Dirac fermions in graphene, *Nature*, Vol. 438, pp. 197-200, 2005.
[3] A.K. Geim and K.S. Novoselov, The rise of graphene, *Nat. Mater.*, Vol. 6, pp. 183-191, 2007.
[4] A.H.C. Neto, F. Guinea, N.M.R. Peres, K.S. Novoselov, and A.K. Geim, The electronic properties of graphene, *Rev. Mod. Phys.*, Vol. 81, p. 109, 2009.
[5] M. Xu, T. Liang, M. Shi, and H. Chen, Graphene-like two-dimensional materials, *Chem. Rev.*, Vol. 113, pp. 3766-3798, 2013.

[6] S.Z. Butler, S.M. Hollen, L. Cao, Y. Cui, J.A. Gupta, H.R. Gutierrez, T.F. Heinz, S.S. Hong, J. Huang, and A.F. Ismach, Progress, challenges, and opportunities in two-dimensional materials beyond graphene, *ACS Nano*, Vol. 7, pp. 2898-2926, 2013.

[7] Q.H. Wang, K. Kalantar-Zadeh, A. Kis, J.N. Coleman, and M.S. Strano, Electronics and optoelectronics of two-dimensional transition metal dichalcogenides, *Nat. Nanotechnol.*, Vol. 7, pp. 699-712, 2012.

[8] M. Chhowalla, H.S. Shin, G. Eda, L.-J. Li, K.P. Loh, and H. Zhang, The chemistry of two-dimensional layered transition metal dichalcogenide nanosheets, *Nat. Chem.*, Vol. 5, pp. 263-275, 2013.

[9] D. Li, Y. Wu, P. Kim, L. Shi, P. Yang, and A. Majumdar, Thermal conductivity of individual silicon nanowires, *Appl. Phys. Lett.*, Vol. 83, pp. 2934-2936, 2003.

[10] S.-M. Lee, D.G. Cahill, and R. Venkatasubramanian, Thermal conductivity of Si–Ge superlattices, *Appl. Phys. Lett.*, Vol. 70, pp. 2957-2959, 1997.

[11] R. Yang and G. Chen, Thermal conductivity modeling of periodic two-dimensional nanocomposites, *Phys. Rev. B*, Vol. 69, p. 195316, 2004.

[12] M.-S. Jeng, R. Yang, D. Song, and G. Chen, Modeling the thermal conductivity and phonon transport in nanoparticle composites using Monte Carlo simulation, *J. Heat Transfer*, Vol. 130, p. 042410, 2008.

[13] M.N. Luckyanova, J. Garg, K. Esfarjani, A. Jandl, M.T. Bulsara, A.J. Schmidt, A.J. Minnich, S. Chen, M.S. Dresselhaus, and Z. Ren, Coherent phonon heat conduction in superlattices, *Science*, Vol. 338, pp. 936-939, 2012.

[14] M.S. Dresselhaus, G. Chen, M.Y. Tang, R. Yang, H. Lee, D. Wang, Z. Ren, J.P. Fleurial, and P. Gogna, New Directions for Low‐Dimensional Thermoelectric Materials, *Adv. Mater.*, Vol. 19, pp. 1043-1053, 2007.

[15] W. Kim, R. Wang, and A. Majumdar, Nanostructuring expands thermal limits, *Nano Today*, Vol. 2, pp. 40-47, 2007.

[16] D.R. Clarke, M. Oechsner, and N.P. Padture, Thermal-barrier coatings for more efficient gas-turbine engines, *MRS Bull.*, Vol. 37, pp. 891-898, 2012.

[17] A.A. Balandin, Thermal properties of graphene and nanostructured carbon materials, *Nat. Mater.*, Vol. 10, pp. 569-581, 2011.

[18] M.M. Sadeghi, M.T. Pettes, and L. Shi, Thermal transport in graphene, *Solid State Commun.*, Vol. 152, pp. 1321-1330, 2012.

[19] E. Pop, V. Varshney, and A.K. Roy, Thermal properties of graphene: Fundamentals and applications, *MRS Bull.*, Vol. 37, pp. 1273-1281, 2012.

[20] D.L. Nika and A.A. Balandin, Two-dimensional phonon transport in graphene, *J. Phys.: Condens. Matter*, Vol. 24, p. 233203, 2012.

[21] C. Dames, Measuring the thermal conductivity of thin films: 3 omega and related electrothermal methods, In G. Chen, V. Prasad, and Y. Jaluria, Eds., *Annual Review of Heat Transfer*, New York, Begell House Publishers, 2013.

[22] D.G. Cahill, Thermal conductivity measurement from 30 to 750 K: the 3ω method, *Rev. Sci. Instrum.*, Vol. 61, pp. 802-808, 1990.

[23] D.G. Cahill, Analysis of heat flow in layered structures for time-domain thermoreflectance, *Rev. Sci. Instrum.*, Vol. 75, pp. 5119-5122, 2004.

[24] A.J. Schmidt, Pump-probe thermoreflectance, In G. Chen, V. Prasad, and Y. Jaluria, Eds., *Annual Review of Heat Transfer*, New York, Begell House Publishers, 2013.

[25] A.A. Balandin, S. Ghosh, W. Bao, I. Calizo, D. Teweldebrhan, F. Miao, and C.N. Lau, Superior thermal conductivity of single-layer graphene, *Nano Lett.*, Vol. 8, pp. 902-907, 2008.
[26] M.T. Pettes, I. Jo, Z. Yao, and L. Shi, Influence of polymeric residue on the thermal conductivity of suspended bilayer graphene, *Nano Lett.*, Vol. 11, pp. 1195-1200, 2011.
[27] S. Ghosh, W. Bao, D.L. Nika, S. Subrina, E.P. Pokatilov, C.N. Lau, and A.A. Balandin, Dimensional crossover of thermal transport in few-layer graphene, *Nat. Mater.*, Vol. 9, pp. 555-558, 2010.
[28] S. Chen, Q. Wu, C. Mishra, J. Kang, H. Zhang, K. Cho, W. Cai, A.A. Balandin, and R.S. Ruoff, Thermal conductivity of isotopically modified graphene, *Nat. Mater.*, Vol. 11, pp. 203-207, 2012.
[29] S. Sahoo, A.P. Gaur, M. Ahmadi, M.J.-F. Guinel, and R.S. Katiyar, Temperature-dependent Raman studies and thermal conductivity of few-layer $MoS_2$, *J. Phys. Chem. C*, Vol. 117, pp. 9042-9047, 2013.
[30] W. Cai, A.L. Moore, Y. Zhu, X. Li, S. Chen, L. Shi, and R.S. Ruoff, Thermal transport in suspended and supported monolayer graphene grown by chemical vapor deposition, *Nano Lett.*, Vol. 10, pp. 1645-1651, 2010.
[31] J.-U. Lee, D. Yoon, H. Kim, S.W. Lee, and H. Cheong, Thermal conductivity of suspended pristine graphene measured by Raman spectroscopy, *Phys. Rev. B*, Vol. 83, p. 081419, 2011.
[32] C. Faugeras, B. Faugeras, M. Orlita, M. Potemski, R.R. Nair, and A. Geim, Thermal conductivity of graphene in corbino membrane geometry, *ACS Nano*, Vol. 4, pp. 1889-1892, 2010.
[33] S. Chen, A.L. Moore, W. Cai, J.W. Suk, J. An, C. Mishra, C. Amos, C.W. Magnuson, J. Kang, and L. Shi, Raman measurements of thermal transport in suspended monolayer graphene of variable sizes in vacuum and gaseous environments, *ACS Nano*, Vol. 5, pp. 321-328, 2010.
[34] J.H. Seol, I. Jo, A.L. Moore, L. Lindsay, Z.H. Aitken, M.T. Pettes, X. Li, Z. Yao, R. Huang, and D. Broido, Two-dimensional phonon transport in supported graphene, *Science*, Vol. 328, pp. 213-216, 2010.
[35] J.S. Reparaz, E. Chavez-Angel, M.R. Wagner, B. Graczykowski, J. Gomis-Bresco, F. Alzina, and C.M.S. Torres, A novel contactless technique for thermal field mapping and thermal conductivity determination: Two-Laser Raman Thermometry, *Rev. Sci. Instrum.*, Vol. 85, p. 034901, 2014.
[36] M.E. Siemens, Q. Li, R. Yang, K.A. Nelson, E.H. Anderson, M.M. Murnane, and H.C. Kapteyn, Quasi-ballistic thermal transport from nanoscale interfaces observed using ultrafast coherent soft X-ray beams, *Nat. Mater.*, Vol. 9, pp. 26-30, 2010.
[37] L. Shi, D. Li, C. Yu, W. Jang, D. Kim, Z. Yao, P. Kim, and A. Majumdar, Measuring thermal and thermoelectric properties of one-dimensional nanostructures using a microfabricated device, *J. Heat Transfer*, Vol. 125, pp. 881-888, 2003.
[38] P. Kim, L. Shi, A. Majumdar, and P. McEuen, Thermal transport measurements of individual multiwalled nanotubes, *Phys. Rev. Lett.*, Vol. 87, p. 215502, 2001.
[39] I. Jo, M.T. Pettes, J. Kim, K. Watanabe, T. Taniguchi, Z. Yao, and L. Shi, Thermal conductivity and phonon transport in suspended few-layer hexagonal boron nitride, *Nano Lett.*, Vol. 13, pp. 550-554, 2013.
[40] D. Liu, R. Xie, N. Yang, B. Li, and J.T. Thong, Profiling nanowire thermal resistance with a spatial resolution of nanometers, *Nano Lett.*, Vol. 14, pp. 806-812, 2014.
[41] Z. Wang, R. Xie, C.T. Bui, D. Liu, X. Ni, B. Li, and J.T. Thong, Thermal transport in suspended and supported few-layer graphene, *Nano Lett.*, Vol. 11, pp. 113-118, 2010.

[42] S.G. Volz and G. Chen, Molecular-dynamics simulation of thermal conductivity of silicon crystals, *Phys. Rev. B*, Vol. 61, p. 2651, 2000.
[43] P.K. Schelling, S.R. Phillpot, and P. Keblinski, Comparison of atomic-level simulation methods for computing thermal conductivity, *Phys. Rev. B*, Vol. 65, p. 144306, 2002.
[44] D.A. Broido, A. Ward, and N. Mingo, Lattice thermal conductivity of silicon from empirical interatomic potentials, *Phys. Rev. B*, Vol. 72, p. 014308, 2005.
[45] K. Esfarjani, G. Chen, and H.T. Stokes, Heat transport in silicon from first-principles calculations, *Phys. Rev. B*, Vol. 84, p. 085204, 2011.
[46] N. Mingo and L. Yang, Phonon transport in nanowires coated with an amorphous material: An atomistic Green's function approach, *Phys. Rev. B*, Vol. 68, p. 245406, 2003.
[47] W. Zhang, T.S. Fisher, and N. Mingo, The atomistic Green's function method: An efficient simulation approach for nanoscale phonon transport, *Numer. Heat Transfer, Part B*, Vol. 51, pp. 333-349, 2007.
[48] J.-S. Wang, J. Wang, and J.T. Lü, Quantum thermal transport in nanostructures, *Eur. Phys. J. B*, Vol. 62, pp. 381-404, 2008.
[49] X. Zhang, H. Xie, M. Hu, H. Bao, S. Yue, G. Qin, and G. Su, Thermal conductivity of silicene calculated using an optimized Stillinger-Weber potential, *Phys. Rev. B*, Vol. 89, p. 054310, 2014.
[50] J.-W. Jiang, H.S. Park, and T. Rabczuk, Molecular dynamics simulations of single-layer molybdenum disulphide ($MoS_2$): Stillinger-Weber parametrization, mechanical properties, and thermal conductivity, *J. Appl. Phys.*, Vol. 114, p. 064307, 2013.
[51] S. Baroni, S. De Gironcoli, A. Dal Corso, and P. Giannozzi, Phonons and related crystal properties from density-functional perturbation theory, *Rev. Mod. Phys.*, Vol. 73, p. 515, 2001.
[52] K. Esfarjani and H.T. Stokes, Method to extract anharmonic force constants from first principles calculations, *Phys. Rev. B*, Vol. 77, p. 144112, 2008.
[53] X. Tang and B. Fultz, First-principles study of phonon linewidths in noble metals, *Phys. Rev. B*, Vol. 84, p. 054303, 2011.
[54] X. Gu and R. Yang, First-principles prediction of phononic thermal conductivity of silicene: A comparison with graphene, *J. Appl. Phys.*, Vol. 117, p. 025102, 2015.
[55] X. Gu and R. Yang, Phonon transport in single-layer transition metal dichalcogenides: A first-principles study, *Appl. Phys. Lett.*, Vol. 105, p. 131903, 2014.
[56] G.P. Srivastava, *The physics of phonons*, Adam Hilger, Bristol, Philadelphia and New York, 1990
[57] G. Chen, *Nanoscale energy transport and conversion: a parallel treatment of electrons, molecules, phonons, and photons*, Oxford University Press, USA, 2005
[58] B.D. Kong, S. Paul, M.B. Nardelli, and K.W. Kim, First-principles analysis of lattice thermal conductivity in monolayer and bilayer graphene, *Phys. Rev. B*, Vol. 80, p. 033406, 2009.
[59] Y. Cai, J. Lan, G. Zhang, and Y.-W. Zhang, Lattice vibrational modes and phonon thermal conductivity of monolayer $MoS_2$, *Phys. Rev. B*, Vol. 89, p. 035438, 2014.
[60] P.G. Klemens and D.F. Pedraza, Thermal conductivity of graphite in the basal plane, *Carbon*, Vol. 32, pp. 735-741, 1994.
[61] P.G. Klemens, Thermal conductivity and lattice vibrational modes, In H. Huntington, F. Seitz, and D. Turnbull, Eds., *Solid state physics*, New York, Academic, pp. 1-98, 1958.
[62] L. Lindsay, D.A. Broido, and N. Mingo, Flexural phonons and thermal transport in graphene, *Phys. Rev. B*, Vol. 82, p. 115427, 2010.

[63] M. Omini and A. Sparavigna, Beyond the isotropic-model approximation in the theory of thermal conductivity, *Phys. Rev. B*, Vol. 53, p. 9064, 1996.
[64] D.A. Broido, M. Malorny, G. Birner, N. Mingo, and D.A. Stewart, Intrinsic lattice thermal conductivity of semiconductors from first principles, *Appl. Phys. Lett.*, Vol. 91, p. 231922, 2007.
[65] J. Garg, N. Bonini, B. Kozinsky, and N. Marzari, Role of disorder and anharmonicity in the thermal conductivity of silicon-germanium alloys: A first-principles study, *Phys. Rev. Lett.*, Vol. 106, p. 045901, 2011.
[66] L. Lindsay, W. Li, J. Carrete, N. Mingo, D. Broido, and T. Reinecke, Phonon thermal transport in strained and unstrained graphene from first principles, *Phys. Rev. B*, Vol. 89, p. 155426, 2014.
[67] W. Li, J. Carrete, and N. Mingo, Thermal conductivity and phonon linewidths of monolayer $MoS_2$ from first principles, *Appl. Phys. Lett.*, Vol. 103, p. 253103, 2013.
[68] J. Shiomi, Nonequilibrium molecular dynamics methods for heat conduction calculations, In G. Chen, V. Prasad, and Y. Jaluria, Eds., *Annual Review of Heat Transfer*, New York, Begell House Publishers, pp. 177-203, 2014.
[69] A.J. McGaughey and J.M. Larkin, Predicting phonon properties from equilibrium molecular dynamics simulations, In G. Chen, V. Prasad, and Y. Jaluria, Eds., *Annual Review of Heat Transfer*, New York, Begell House Publishers, 2014.
[70] J.M. Dickey and A. Paskin, Computer simulation of the lattice dynamics of solids, *Phys. Rev.*, Vol. 188, p. 1407, 1969.
[71] A.J. Ladd, B. Moran, and W.G. Hoover, Lattice thermal conductivity: A comparison of molecular dynamics and anharmonic lattice dynamics, *Phys. Rev. B*, Vol. 34, p. 5058, 1986.
[72] P.K. Schelling, S.R. Phillpot, and P. Keblinski, Phonon wave-packet dynamics at semiconductor interfaces by molecular-dynamics simulation, *Appl. Phys. Lett.*, Vol. 80, pp. 2484-2486, 2002.
[73] J.E. Turney, A.J.H. McGaughey, and C.H. Amon, Assessing the applicability of quantum corrections to classical thermal conductivity predictions, *Phys. Rev. B*, Vol. 79, p. 224305, 2009.
[74] L. Lindsay and D.A. Broido, Optimized Tersoff and Brenner empirical potential parameters for lattice dynamics and phonon thermal transport in carbon nanotubes and graphene, *Phys. Rev. B*, Vol. 81, p. 205441, 2010.
[75] J. Shiomi, K. Esfarjani, and G. Chen, Thermal conductivity of half-Heusler compounds from first-principles calculations, *Phys. Rev. B*, Vol. 84, p. 104302, 2011.
[76] X. Li, K. Maute, M.L. Dunn, and R. Yang, Strain effects on the thermal conductivity of nanostructures, *Phys. Rev. B*, Vol. 81, p. 245318, 2010.
[77] X. Li and R. Yang, Effect of lattice mismatch on phonon transmission and interface thermal conductance across dissimilar material interfaces, *Phys. Rev. B*, Vol. 86, p. 054305, 2012.
[78] X. Li and R. Yang, Size-dependent phonon transmission across dissimilar material interfaces, *J. Phys.: Condens. Matter*, Vol. 24, p. 155302, 2012.
[79] L.G.C. Rego and G. Kirczenow, Quantized thermal conductance of dielectric quantum wires, *Phys. Rev. Lett.*, Vol. 81, p. 232, 1998.
[80] N. Katcho, J. Carrete, W. Li, and N. Mingo, Effect of nitrogen and vacancy defects on the thermal conductivity of diamond: An ab initio Green's function approach, *Phys. Rev. B*, Vol. 90, p. 094117, 2014.
[81] A. Kundu, N. Mingo, D. Broido, and D. Stewart, Role of light and heavy embedded nanoparticles on the thermal conductivity of SiGe alloys, *Phys. Rev. B*, Vol. 84, p. 125426, 2011.

[82] P. Chen, N. Katcho, J. Feser, W. Li, M. Glaser, O. Schmidt, D.G. Cahill, N. Mingo, and A. Rastelli, Role of surface-segregation-driven intermixing on the thermal transport through planar si/ge superlattices, *Phys. Rev. Lett.*, Vol. 111, p. 115901, 2013.

[83] S. Ghosh, I. Calizo, D. Teweldebrhan, E. Pokatilov, D. Nika, A. Balandin, W. Bao, F. Miao, and C.N. Lau, Extremely high thermal conductivity of graphene: Prospects for thermal management applications in nanoelectronic circuits, *Appl. Phys. Lett.*, Vol. 92, pp. 151911-151911-151913, 2008.

[84] R.R. Nair, P. Blake, A.N. Grigorenko, K.S. Novoselov, T.J. Booth, T. Stauber, N.M.R. Peres, and A.K. Geim, Fine structure constant defines visual transparency of graphene, *Science*, Vol. 320, pp. 1308-1308, 2008.

[85] K.F. Mak, M.Y. Sfeir, Y. Wu, C.H. Lui, J.A. Misewich, and T.F. Heinz, Measurement of the optical conductivity of graphene, *Phys. Rev. Lett.*, Vol. 101, p. 196405, 2008.

[86] A. Kuzmenko, E. Van Heumen, F. Carbone, and D. Van Der Marel, Universal optical conductance of graphite, *Phys. Rev. Lett.*, Vol. 100, p. 117401, 2008.

[87] T. Stauber, N.M.R. Peres, and A.K. Geim, Optical conductivity of graphene in the visible region of the spectrum, *Phys. Rev. B*, Vol. 78, p. 085432, 2008.

[88] X. Xu, L.F. Pereira, Y. Wang, J. Wu, K. Zhang, X. Zhao, S. Bae, C.T. Bui, R. Xie, and J.T. Thong, Length-dependent thermal conductivity in suspended single-layer graphene, *Nature Commun.*, Vol. 5, 2014.

[89] D.L. Nika, E.P. Pokatilov, A.S. Askerov, and A.A. Balandin, Phonon thermal conduction in graphene: Role of Umklapp and edge roughness scattering, *Phys. Rev. B*, Vol. 79, p. 155413, 2009.

[90] D. Singh, J.Y. Murthy, and T.S. Fisher, Mechanism of thermal conductivity reduction in few-layer graphene, *J. Appl. Phys.*, Vol. 110, p. 044317, 2011.

[91] G. Fugallo, A. Cepellotti, L. Paulatto, M. Lazzeri, N. Marzari, and F. Mauri, Thermal Conductivity of Graphene and Graphite: Collective Excitations and Mean Free Paths, *Nano Lett.*, Vol. 14, pp. 6109-6114, 2014.

[92] N. Bonini, J. Garg, and N. Marzari, Acoustic phonon lifetimes and thermal transport in free-standing and strained graphene, *Nano Lett.*, Vol. 12, pp. 2673-2678, 2012.

[93] I. Vlassiouk, S. Smirnov, I. Ivanov, P.F. Fulvio, S. Dai, H. Meyer, M. Chi, D. Hensley, P. Datskos, and N.V. Lavrik, Electrical and thermal conductivity of low temperature CVD graphene: the effect of disorder, *Nanotechnology*, Vol. 22, p. 275716, 2011.

[94] S. Chen, Q. Li, Q. Zhang, Y. Qu, H. Ji, R.S. Ruoff, and W. Cai, Thermal conductivity measurements of suspended graphene with and without wrinkles by micro-Raman mapping, *Nanotechnology*, Vol. 23, p. 365701, 2012.

[95] H. Li, H. Ying, X. Chen, D.L. Nika, A.I. Cocemasov, W. Cai, A.A. Balandin, and S. Chen, Thermal conductivity of twisted bilayer graphene, *Nanoscale*, Vol. 6, pp. 13402-13408, 2014.

[96] V.E. Dorgan, A. Behnam, H.J. Conley, K.I. Bolotin, and E. Pop, High-field electrical and thermal transport in suspended graphene, *Nano Lett.*, Vol. 13, pp. 4581-4586, 2013.

[97] J. Wang, L. Zhu, J. Chen, B. Li, and J.T. Thong, Suppressing Thermal Conductivity of Suspended Tri‐layer Graphene by Gold Deposition, *Adv. Mater.*, Vol. 25, pp. 6884-6888, 2013.

[98] Z. Aksamija and I. Knezevic, Lattice thermal conductivity of graphene nanoribbons: anisotropy and edge roughness scattering, *Appl. Phys. Lett.*, Vol. 98, p. 141919, 2011.

[99] Z. Aksamija and I. Knezevic, Lattice thermal transport in large-area polycrystalline graphene, *Phys. Rev. B*, Vol. 90, p. 035419, 2014.

[100] S. Mei, L. Maurer, Z. Aksamija, and I. Knezevic, Full-dispersion Monte Carlo simulation of phonon transport in micron-sized graphene nanoribbons, *J. Appl. Phys.*, Vol. 116, p. 164307, 2014.

[101] W.J. Evans, L. Hu, and P. Keblinski, Thermal conductivity of graphene ribbons from equilibrium molecular dynamics: Effect of ribbon width, edge roughness, and hydrogen termination, *Appl. Phys. Lett.*, Vol. 96, p. 203112, 2010.

[102] Z. Guo, D. Zhang, and X.-G. Gong, Thermal conductivity of graphene nanoribbons, *Appl. Phys. Lett.*, Vol. 95, p. 163103, 2009.

[103] L.F.C. Pereira and D. Donadio, Divergence of the thermal conductivity in uniaxially strained graphene, *Phys. Rev. B*, Vol. 87, p. 125424, 2013.

[104] H. Zhang, G. Lee, and K. Cho, Thermal transport in graphene and effects of vacancy defects, *Phys. Rev. B*, Vol. 84, p. 115460, 2011.

[105] J. Haskins, A. Kınacı, C. Sevik, H. Sevinçli, G. Cuniberti, and T. Çağın, Control of thermal and electronic transport in defect-engineered graphene nanoribbons, *ACS Nano*, Vol. 5, pp. 3779-3787, 2011.

[106] N. Mingo, K. Esfarjani, D. Broido, and D. Stewart, Cluster scattering effects on phonon conduction in graphene, *Phys. Rev. B*, Vol. 81, p. 045408, 2010.

[107] B. Mortazavi, M. Pötschke, and G. Cuniberti, Multiscale modeling of thermal conductivity of polycrystalline graphene sheets, *Nanoscale*, Vol. 6, pp. 3344-3352, 2014.

[108] H. Liu, Y. Lin, and S.-N. Luo, Grain Boundary Energy and Grain Size Dependences of Thermal Conductivity of Polycrystalline Graphene, *J. Phys. Chem. C*, Vol. 118, pp. 24797-24802, 2014.

[109] C.R. Dean, A.F. Young, I. Meric, C. Lee, L. Wang, S. Sorgenfrei, K. Watanabe, T. Taniguchi, P. Kim, and K.L. Shepard, Boron nitride substrates for high-quality graphene electronics, *Nat. Nanotechnol.*, Vol. 5, pp. 722-726, 2010.

[110] C. Sevik, A. Kinaci, J.B. Haskins, and T. Çağın, Characterization of thermal transport in low-dimensional boron nitride nanostructures, *Phys. Rev. B*, Vol. 84, p. 085409, 2011.

[111] L. Lindsay and D.A. Broido, Enhanced thermal conductivity and isotope effect in single-layer hexagonal boron nitride, *Phys. Rev. B*, Vol. 84, p. 155421, 2011.

[112] L. Lindsay and D.A. Broido, Theory of thermal transport in multilayer hexagonal boron nitride and nanotubes, *Phys. Rev. B*, Vol. 85, p. 035436, 2012.

[113] E.K. Sichel, R.E. Miller, M.S. Abrahams, and C.J. Buiocchi, Heat capacity and thermal conductivity of hexagonal pyrolytic boron nitride, *Phys. Rev. B*, Vol. 13, p. 4607, 1976.

[114] H. Zhou, J. Zhu, Z. Liu, Z. Yan, X. Fan, J. Lin, G. Wang, Q. Yan, T. Yu, and P.M. Ajayan, High thermal conductivity of suspended few-layer hexagonal boron nitride sheets, *Nano Res.*, Vol. 7, pp. 1232-1240, 2014.

[115] T. Ouyang, Y. Chen, Y. Xie, K. Yang, Z. Bao, and J. Zhong, Thermal transport in hexagonal boron nitride nanoribbons, *Nanotechnology*, Vol. 21, p. 245701, 2010.

[116] K. Yang, Y. Chen, Y. Xie, X. Wei, T. Ouyang, and J. Zhong, Effect of triangle vacancy on thermal transport in boron nitride nanoribbons, *Solid State Commun.*, Vol. 151, pp. 460-464, 2011.

[117] K. Takeda and K. Shiraishi, Theoretical possibility of stage corrugation in Si and Ge analogs of graphite, *Phys. Rev. B*, Vol. 50, p. 14916, 1994.

[118] A. Fleurence, R. Friedlein, T. Ozaki, H. Kawai, Y. Wang, and Y. Yamada-Takamura, Experimental evidence for epitaxial silicene on diboride thin films, *Phys. Rev. Lett.*, Vol. 108, p. 245501, 2012.

[119] P. Vogt, P. De Padova, C. Quaresima, J. Avila, E. Frantzeskakis, M.C. Asensio, A. Resta, B. Ealet, and G. Le Lay, Silicene: compelling experimental evidence for graphenelike two-dimensional silicon, *Phys. Rev. Lett.*, Vol. 108, p. 155501, 2012.
[120] L. Chen, C.-C. Liu, B. Feng, X. He, P. Cheng, Z. Ding, S. Meng, Y. Yao, and K. Wu, Evidence for Dirac fermions in a honeycomb lattice based on silicon, *Phys. Rev. Lett.*, Vol. 109, p. 056804, 2012.
[121] N. Drummond, V. Zolyomi, and V. Fal'Ko, Electrically tunable band gap in silicene, *Phys. Rev. B*, Vol. 85, p. 075423, 2012.
[122] Z. Ni, Q. Liu, K. Tang, J. Zheng, J. Zhou, R. Qin, Z. Gao, D. Yu, and J. Lu, Tunable bandgap in silicene and germanene, *Nano Lett.*, Vol. 12, pp. 113-118, 2011.
[123] H. Xie, M. Hu, and H. Bao, Thermal conductivity of silicene from first-principles, *Appl. Phys. Lett.*, Vol. 104, p. 131906, 2014.
[124] J. Tersoff, New empirical approach for the structure and energy of covalent systems, *Phys. Rev. B*, Vol. 37, p. 6991, 1988.
[125] M.I. Baskes, Modified embedded-atom potentials for cubic materials and impurities, *Phys. Rev. B*, Vol. 46, p. 2727, 1992.
[126] Q.-X. Pei, Y.-W. Zhang, Z.-D. Sha, and V.B. Shenoy, Tuning the thermal conductivity of silicene with tensile strain and isotopic doping: A molecular dynamics study, *J. Appl. Phys.*, Vol. 114, p. 033526, 2013.
[127] M. Hu, X. Zhang, and D. Poulikakos, Anomalous thermal response of silicene to uniaxial stretching, *Phys. Rev. B*, Vol. 87, p. 195417, 2013.
[128] H.-p. Li and R.-Q. Zhang, Vacancy-defect–induced diminution of thermal conductivity in silicene, *EPL (Europhysics Letters)*, Vol. 99, p. 36001, 2012.
[129] L. Wang and H. Sun, Thermal conductivity of silicon and carbon hybrid monolayers: a molecular dynamics study, *J. Mol. Model.*, Vol. 18, pp. 4811-4818, 2012.
[130] L. Pan, H.J. Liu, X.J. Tan, H.Y. Lv, J. Shi, X.F. Tang, and G. Zheng, Thermoelectric properties of armchair and zigzag silicene nanoribbons, *Phys. Chem. Chem. Phys.*, Vol. 14, pp. 13588-13593, 2012.
[131] B. Radisavljevic, A. Radenovic, J. Brivio, V. Giacometti, and A. Kis, Single-layer $MoS_2$ transistors, *Nat. Nanotechnol.*, Vol. 6, pp. 147-150, 2011.
[132] R.S. Sundaram, M. Engel, A. Lombardo, R. Krupke, A.C. Ferrari, P. Avouris, and M. Steiner, Electroluminescence in single layer $MoS_2$, *Nano Lett.*, Vol. 13, pp. 1416-1421, 2013.
[133] O. Lopez-Sanchez, D. Lembke, M. Kayci, A. Radenovic, and A. Kis, Ultrasensitive photodetectors based on monolayer $MoS_2$, *Nat. Nanotechnol.*, Vol. 8, pp. 497-501, 2013.
[134] T. Liang, S.R. Phillpot, and S.B. Sinnott, Parametrization of a reactive many-body potential for Mo–S systems, *Phys. Rev. B*, Vol. 79, p. 245110, 2009.
[135] V. Varshney, S.S. Patnaik, C. Muratore, A.K. Roy, A.A. Voevodin, and B.L. Farmer, MD simulations of molybdenum disulphide ($MoS_2$): Force-field parameterization and thermal transport behavior, *Comput. Mater. Sci.*, Vol. 48, pp. 101-108, 2010.
[136] T. Onodera, Y. Morita, A. Suzuki, M. Koyama, H. Tsuboi, N. Hatakeyama, A. Endou, H. Takaba, M. Kubo, and F. Dassenoy, A computational chemistry study on friction of h-$MoS_2$. Part I. Mechanism of single sheet lubrication, *J. Phys. Chem. B*, Vol. 113, pp. 16526-16536, 2009.
[137] R. Yan, J.R. Simpson, S. Bertolazzi, J. Brivio, M. Watson, X. Wu, A. Kis, T. Luo, A.R. Hight Walker, and H.G. Xing, Thermal conductivity of monolayer molybdenum disulfide obtained from temperature-dependent Raman spectroscopy, *ACS Nano*, Vol. 8, pp. 986-993, 2014.

[138] I. Jo, M.T. Pettes, E. Ou, W. Wu, and L. Shi, Basal-plane thermal conductivity of few-layer molybdenum disulfide, *Appl. Phys. Lett.*, Vol. 104, p. 201902, 2014.

[139] A. Taube, A. Lapinska, J. Judek, and M. Zdrojek, Temperature-dependent thermal properties of supported $MoS_2$ monolayers, *ACS Appl. Mater. Interfaces*, 2015.

[140] J. Liu, G.-M. Choi, and D.G. Cahill, Measurement of the anisotropic thermal conductivity of molybdenum disulfide by the time-resolved magneto-optic Kerr effect, *J. Appl. Phys.*, Vol. 116, p. 233107, 2014.

[141] N. Peimyoo, J. Shang, W. Yang, Y. Wang, C. Cong, and T. Yu, Thermal conductivity determination of suspended mono-and bilayer $WS_2$ by Raman spectroscopy, *Nano Res.*, pp. 1-12, 2014.

[142] X. Liu, G. Zhang, Q.-X. Pei, and Y.-W. Zhang, Phonon thermal conductivity of monolayer $MoS_2$ sheet and nanoribbons, *Appl. Phys. Lett.*, Vol. 103, p. 133113, 2013.

[143] C. Muratore, V. Varshney, J.J. Gengler, J. Hu, J.E. Bultman, A.K. Roy, B.L. Farmer, and A.A. Voevodin, Thermal anisotropy in nano-crystalline $MoS_2$ thin films, *Phys. Chem. Chem. Phys.*, Vol. 16, pp. 1008-1014, 2014.

[144] D. Teweldebrhan, V. Goyal, M. Rahman, and A.A. Balandin, Atomically-thin crystalline films and ribbons of bismuth telluride, *Appl. Phys. Lett.*, Vol. 96, p. 3107, 2010.

[145] V. Goyal, D. Teweldebrhan, and A. Balandin, Mechanically-exfoliated stacks of thin films of $Bi_2Te_3$ topological insulators with enhanced thermoelectric performance, *Appl. Phys. Lett.*, Vol. 97, p. 133117, 2010.

[146] D. Teweldebrhan, V. Goyal, and A.A. Balandin, Exfoliation and characterization of bismuth telluride atomic quintuples and quasi-two-dimensional crystals, *Nano Lett.*, Vol. 10, pp. 1209-1218, 2010.

[147] B. Qiu and X. Ruan, Thermal conductivity prediction and analysis of few-quintuple $Bi_2Te_3$ thin films: A molecular dynamics study, *Appl. Phys. Lett.*, Vol. 97, p. 183107, 2010.

[148] L. Zhu, G. Zhang, and B. Li, Coexistence of size-dependent and size-independent thermal conductivities in phosphorene, *Phys. Rev. B*, Vol. 90, p. 214302, 2014.

[149] A. Jain and A.J.H. McGaughey, Strongly anisotropic in-plane thermal transport in single-layer black phosphorene, *Sci. Rep.*, Vol. 5, 2015.

[150] G. Qin, Q.-B. Yan, Z. Qin, S.-Y. Yue, M. Hu, and G. Su, Anisotropic intrinsic lattice thermal conductivity of phosphorene from first principles, *Phys. Chem. Chem. Phys.*, 2015.

[151] Z. Luo, J. Maassen, Y. Deng, Y. Du, M.S. Lundstrom, P.D. Ye, and X. Xu, Anisotropic in-plane thermal conductivity observed in few-layer black phosphorus, *arXiv preprint arXiv:1503.06167*, 2015.

[152] G. Barbarino, C. Melis, and L. Colombo, Intrinsic thermal conductivity in monolayer graphene is ultimately upper limited: A direct estimation by atomistic simulations, *Phys. Rev. B*, Vol. 91, p. 035416, 2015.

[153] M.-H. Bae, Z. Li, Z. Aksamija, P.N. Martin, F. Xiong, Z.-Y. Ong, I. Knezevic, and E. Pop, Ballistic to diffusive crossover of heat flow in graphene ribbons, *Nature Commun.*, Vol. 4, p. 1734, 2013.

[154] X. Mu, X. Wu, T. Zhang, D.B. Go, and T. Luo, Thermal transport in graphene oxide–from ballistic extreme to amorphous limit, *Sci. Rep.*, Vol. 4, 2014.

[155] M. Park, S.-C. Lee, and Y.-S. Kim, Length-dependent lattice thermal conductivity of graphene and its macroscopic limit, *J. Appl. Phys.*, Vol. 114, p. 053506, 2013.

[156] N. Yang, X. Xu, G. Zhang, and B. Li, Thermal transport in nanostructures, *AIP Adv.*, Vol. 2, p. 041410, 2012.

[157] G. Fugallo, M. Lazzeri, L. Paulatto, and F. Mauri, Ab initio variational approach for evaluating lattice thermal conductivity, *Phys. Rev. B*, Vol. 88, p. 045430, 2013.
[158] L.R. Lindsay, *Theory of phonon thermal transport in single-walled carbon nanotubes and graphene*, Ph.D. Thesis, Boston College, 2010.
[159] X. Li, X. Wang, L. Zhang, S. Lee, and H. Dai, Chemically derived, ultrasmooth graphene nanoribbon semiconductors, *Science*, Vol. 319, pp. 1229-1232, 2008.
[160] J. Cai, P. Ruffieux, R. Jaafar, M. Bieri, T. Braun, S. Blankenburg, M. Muoth, A.P. Seitsonen, M. Saleh, and X. Feng, Atomically precise bottom-up fabrication of graphene nanoribbons, *Nature*, Vol. 466, pp. 470-473, 2010.
[161] H. Sevinçli and G. Cuniberti, Enhanced thermoelectric figure of merit in edge-disordered zigzag graphene nanoribbons, *Phys. Rev. B*, Vol. 81, p. 113401, 2010.
[162] J. Hu, X. Ruan, and Y.P. Chen, Thermal conductivity and thermal rectification in graphene nanoribbons: a molecular dynamics study, *Nano Lett.*, Vol. 9, pp. 2730-2735, 2009.
[163] N. Yang, G. Zhang, and B. Li, Thermal rectification in asymmetric graphene ribbons, *Appl. Phys. Lett.*, Vol. 95, p. 033107, 2009.
[164] A.V. Savin, Y.S. Kivshar, and B. Hu, Suppression of thermal conductivity in graphene nanoribbons with rough edges, *Phys. Rev. B*, Vol. 82, p. 195422, 2010.
[165] J. Hu, S. Schiffli, A. Vallabhaneni, X. Ruan, and Y.P. Chen, Tuning the thermal conductivity of graphene nanoribbons by edge passivation and isotope engineering: A molecular dynamics study, *Appl. Phys. Lett.*, Vol. 97, p. 133107, 2010.
[166] Z.W. Tan, J.-S. Wang, and C.K. Gan, First-principles study of heat transport properties of graphene nanoribbons, *Nano Lett.*, Vol. 11, pp. 214-219, 2010.
[167] Y. Xu, X. Chen, B.-L. Gu, and W. Duan, Intrinsic anisotropy of thermal conductance in graphene nanoribbons, *Appl. Phys. Lett.*, Vol. 95, p. 233116, 2009.
[168] Z. Wei, Y. Chen, and C. Dames, Wave packet simulations of phonon boundary scattering at graphene edges, *J. Appl. Phys.*, Vol. 112, p. 024328, 2012.
[169] Y. Wang, B. Qiu, and X. Ruan, Edge effect on thermal transport in graphene nanoribbons: A phonon localization mechanism beyond edge roughness scattering, *Appl. Phys. Lett.*, Vol. 101, p. 013101, 2012.
[170] L. Lindsay, D.A. Broido, and N. Mingo, Flexural phonons and thermal transport in multilayer graphene and graphite, *Phys. Rev. B*, Vol. 83, p. 235428, 2011.
[171] Z. Wei, Z. Ni, K. Bi, M. Chen, and Y. Chen, In-plane lattice thermal conductivities of multilayer graphene films, *Carbon*, Vol. 49, pp. 2653-2658, 2011.
[172] Z. Yan, C. Jiang, T.R. Pope, C.F. Tsang, J.L. Stickney, P. Goli, J. Renteria, T.T. Salguero, and A.A. Balandin, Phonon and thermal properties of exfoliated TaSe2 thin films, *J. Appl. Phys.*, Vol. 114, p. 204301, 2013.
[173] S.-i. Tamura, Isotope scattering of dispersive phonons in Ge, *Phys. Rev. B*, Vol. 27, pp. 858-866, 1983.
[174] P.G. Klemens, The scattering of low-frequency lattice waves by static imperfections, *Proc. Phys. Soc. Lond., Sect. A*, Vol. 68, p. 1113, 1955.
[175] P.G. Klemens, Theory of the a-plane thermal conductivity of graphite, *J. Wide Bandgap Mater.*, Vol. 7, pp. 332-339, 2000.
[176] W. Kim, J. Zide, A. Gossard, D. Klenov, S. Stemmer, A. Shakouri, and A. Majumdar, Thermal conductivity reduction and thermoelectric figure of merit increase by embedding nanoparticles in crystalline semiconductors, *Phys. Rev. Lett.*, Vol. 96, p. 045901, 2006.

[177] J.-W. Jiang, B.-S. Wang, and J.-S. Wang, First principle study of the thermal conductance in graphene nanoribbon with vacancy and substitutional silicon defects, *Appl. Phys. Lett.*, Vol. 98, p. 113114, 2011.
[178] F. Hao, D. Fang, and Z. Xu, Mechanical and thermal transport properties of graphene with defects, *Appl. Phys. Lett.*, Vol. 99, p. 041901, 2011.
[179] C.A. Ratsifaritana and P.G. Klemens, Scattering of phonons by vacancies, *Int. J. Thermophys.*, Vol. 8, pp. 737-750, 1987.
[180] G. Xie, Y. Shen, X. Wei, L. Yang, H. Xiao, J. Zhong, and G. Zhang, A Bond-order Theory on the Phonon Scattering by Vacancies in Two-dimensional Materials, *Sci. Rep.*, Vol. 4, 2014.
[181] Y. Wei, J. Wu, H. Yin, X. Shi, R. Yang, and M. Dresselhaus, The nature of strength enhancement and weakening by pentagon–heptagon defects in graphene, *Nat. Mater.*, Vol. 11, pp. 759-763, 2012.
[182] A. Bagri, S.-P. Kim, R.S. Ruoff, and V.B. Shenoy, Thermal transport across twin grain boundaries in polycrystalline graphene from nonequilibrium molecular dynamics simulations, *Nano Lett.*, Vol. 11, pp. 3917-3921, 2011.
[183] Y. Lu and J. Guo, Thermal transport in grain boundary of graphene by non-equilibrium Green's function approach, *Appl. Phys. Lett.*, Vol. 101, p. 043112, 2012.
[184] A.Y. Serov, Z.-Y. Ong, and E. Pop, Effect of grain boundaries on thermal transport in graphene, *Appl. Phys. Lett.*, Vol. 102, p. 033104, 2013.
[185] A. Cao and J. Qu, Kapitza conductance of symmetric tilt grain boundaries in graphene, *J. Appl. Phys.*, Vol. 111, p. 053529, 2012.
[186] W.S. Yun, S. Han, S.C. Hong, I.G. Kim, and J. Lee, Thickness and strain effects on electronic structures of transition metal dichalcogenides: 2H-$MX_2$ semiconductors (M= Mo, W; X= S, Se, Te), *Phys. Rev. B*, Vol. 85, p. 033305, 2012.
[187] F. Guinea, M.I. Katsnelson, and A.K. Geim, Energy gaps and a zero-field quantum Hall effect in graphene by strain engineering, *Nature Physics*, Vol. 6, pp. 30-33, 2010.
[188] D. Nandwana and E. Ertekin, Ripples, strain, and misfit dislocations: structure of graphene-boron nitride superlattice interfaces, *Nano Lett.*, 2015.
[189] W.-P. Hsieh, B. Chen, J. Li, P. Keblinski, and D.G. Cahill, Pressure tuning of the thermal conductivity of the layered muscovite crystal, *Phys. Rev. B*, Vol. 80, p. 180302, 2009.
[190] B. Wang, J. Wu, X. Gu, H. Yin, Y. Wei, R. Yang, and M. Dresselhaus, Stable planar single-layer hexagonal silicene under tensile strain and its anomalous Poisson's ratio, *Appl. Phys. Lett.*, Vol. 104, p. 081902, 2014.
[191] N. Wei, L. Xu, H.-Q. Wang, and J.-C. Zheng, Strain engineering of thermal conductivity in graphene sheets and nanoribbons: a demonstration of magic flexibility, *Nanotechnology*, Vol. 22, p. 105705, 2011.
[192] Z.-Y. Ong and E. Pop, Effect of substrate modes on thermal transport in supported graphene, *Phys. Rev. B*, Vol. 84, p. 075471, 2011.
[193] B. Qiu and X. Ruan, Reduction of spectral phonon relaxation times from suspended to supported graphene, *Appl. Phys. Lett.*, Vol. 100, p. 193101, 2012.
[194] Z.-X. Guo, J. Ding, and X.-G. Gong, Substrate effects on the thermal conductivity of epitaxial graphene nanoribbons, *Phys. Rev. B*, Vol. 85, p. 235429, 2012.
[195] J. Chen, G. Zhang, and B. Li, Substrate coupling suppresses size dependence of thermal conductivity in supported graphene, *Nanoscale*, Vol. 5, pp. 532-536, 2013.

[196] W. Jang, Z. Chen, W. Bao, C.N. Lau, and C. Dames, Thickness-dependent thermal conductivity of encased graphene and ultrathin graphite, *Nano Lett.*, Vol. 10, pp. 3909-3913, 2010.
[197] M.M. Sadeghi, I. Jo, and L. Shi, Phonon-interface scattering in multilayer graphene on an amorphous support, *Proc. Natl. Acad. Sci. U.S.A.*, Vol. 110, pp. 16321-16326, 2013.
[198] Z. Wei, J. Yang, W. Chen, K. Bi, D. Li, and Y. Chen, Phonon mean free path of graphite along the c-axis, *Appl. Phys. Lett.*, Vol. 104, p. 081903, 2014.
[199] Q. Fu, J. Yang, Y. Chen, D. Li, and D. Xu, Experimental evidence of very long intrinsic phonon mean free path along the c-axis of graphite, *Appl. Phys. Lett.*, Vol. 106, p. 031905, 2015.
[200] X. Duan, C. Wang, J.C. Shaw, R. Cheng, Y. Chen, H. Li, X. Wu, Y. Tang, Q. Zhang, and A. Pan, Lateral epitaxial growth of two-dimensional layered semiconductor heterojunctions, *Nat. Nanotechnol.*, 2014.
[201] Y. Gong, J. Lin, X. Wang, G. Shi, S. Lei, Z. Lin, X. Zou, G. Ye, R. Vajtai, and B.I. Yakobson, Vertical and in-plane heterostructures from $WS_2/MoS_2$ monolayers, *Nat. Mater.*, Vol. 13, pp. 1135-1142, 2014.
[202] T. Zhu and E. Ertekin, Phonon transport on two-dimensional graphene/boron nitride superlattices, *Phys. Rev. B*, Vol. 90, p. 195209, 2014.
[203] A. Kınacı, J.B. Haskins, C. Sevik, and T. Çağın, Thermal conductivity of BN-C nanostructures, *Phys. Rev. B*, Vol. 86, p. 115410, 2012.
[204] Y. Jing, M. Hu, and L. Guo, Thermal conductivity of hybrid graphene/silicon heterostructures, *J. Appl. Phys.*, Vol. 114, p. 153518, 2013.
[205] J. Song and N.V. Medhekar, Thermal transport in lattice-constrained 2D hybrid graphene heterostructures, *J. Phys.: Condens. Matter*, Vol. 25, p. 445007, 2013.
[206] J.-W. Jiang, J.-S. Wang, and B.-S. Wang, Minimum thermal conductance in graphene and boron nitride superlattice, *Appl. Phys. Lett.*, Vol. 99, p. 043109, 2011.
[207] B. Yang and G. Chen, Partially coherent phonon heat conduction in superlattices, *Phys. Rev. B*, Vol. 67, p. 195311, 2003.
[208] M. Simkin and G. Mahan, Minimum thermal conductivity of superlattices, *Phys. Rev. Lett.*, Vol. 84, p. 927, 2000.
[209] H. Sevinçli, W. Li, N. Mingo, G. Cuniberti, and S. Roche, Effects of domains in phonon conduction through hybrid boron nitride and graphene sheets, *Phys. Rev. B*, Vol. 84, p. 205444, 2011.
[210] B. Liu, F. Meng, C.D. Reddy, J.A. Baimova, N. Srikanth, S.V. Dmitriev, and K. Zhou, Thermal transport in a graphene–$MoS_2$ bilayer heterostructure: a molecular dynamics study, *RSC Advances*, Vol. 5, pp. 29193-29200, 2015.
[211] C.-C. Chen, Z. Li, L. Shi, and S.B. Cronin, Thermal interface conductance across a graphene/hexagonal boron nitride heterojunction, *Appl. Phys. Lett.*, Vol. 104, p. 081908, 2014.
[212] J. Zhang, Y. Hong, and Y. Yue, Thermal transport across graphene and single layer hexagonal boron nitride, *J. Appl. Phys.*, Vol. 117, p. 134307, 2015.
[213] B. Liu, J.A. Baimova, C.D. Reddy, S.V. Dmitriev, W.K. Law, X.Q. Feng, and K. Zhou, Interface thermal conductance and rectification in hybrid graphene/silicene monolayer, *Carbon*, Vol. 79, pp. 236-244, 2014.
[214] Q.-X. Pei, Z.-D. Sha, and Y.-W. Zhang, A theoretical analysis of the thermal conductivity of hydrogenated graphene, *Carbon*, Vol. 49, pp. 4752-4759, 2011.
[215] J.Y. Kim, J.-H. Lee, and J.C. Grossman, Thermal transport in functionalized graphene, *ACS Nano*, Vol. 6, pp. 9050-9057, 2012.

[216] B. Liu, C. Reddy, J. Jiang, J.A. Baimova, S.V. Dmitriev, A.A. Nazarov, and K. Zhou, Morphology and in-plane thermal conductivity of hybrid graphene sheets, *Appl. Phys. Lett.*, Vol. 101, p. 211909, 2012.
[217] W. Huang, Q.-X. Pei, Z. Liu, and Y.-W. Zhang, Thermal conductivity of fluorinated graphene: A non-equilibrium molecular dynamics study, *Chem. Phys. Lett.*, Vol. 552, pp. 97-101, 2012.
[218] B. Liu, C. Reddy, J. Jiang, J.A. Baimova, S.V. Dmitriev, A.A. Nazarov, and K. Zhou, Morphology and in-plane thermal conductivity of hybrid graphene sheets, *Applied Physics Letters*, Vol. 101, p. 211909, 2012.
[219] D.G. Cahill, S.K. Watson, and R.O. Pohl, Lower limit to the thermal conductivity of disordered crystals, *Phys. Rev. B*, Vol. 46, p. 6131, 1992.
[220] S.M. Whittingha, *Intercalation chemistry*, Elsevier, 2012
[221] M.S. Dresselhaus and G. Dresselhaus, Intercalation compounds of graphite, *Adv. Phys.*, Vol. 30, pp. 139-326, 1981.
[222] J. Wilson and A. Yoffe, The transition metal dichalcogenides discussion and interpretation of the observed optical, electrical and structural properties, *Adv. Phys.*, Vol. 18, pp. 193-335, 1969.
[223] J.-P. Issi, J. Heremans, and M.S. Dresselhaus, Electronic and lattice contributions to the thermal conductivity of graphite intercalation compounds, *Phys. Rev. B*, Vol. 27, p. 1333, 1983.
[224] R. Clarke and C. Uher, High pressure properties of graphite and its intercalation compounds, *Adv. Phys.*, Vol. 33, pp. 469-566, 1984.
[225] M. Elzinga, D. Morelli, and C. Uher, Thermal transport properties of $SbCl_5$ graphite, *Phys. Rev. B*, Vol. 26, p. 3312, 1982.
[226] C. Wan, Y. Wang, N. Wang, and K. Koumoto, Low-thermal-conductivity $(MS)_{1+x}(TiS_2)_2$ (M= Pb, Bi, Sn) misfit layer compounds for bulk thermoelectric materials, *Materials*, Vol. 3, pp. 2606-2617, 2010.
[227] C. Wan, Y. Wang, W. Norimatsu, M. Kusunoki, and K. Koumoto, Nanoscale stacking faults induced low thermal conductivity in thermoelectric layered metal sulfides, *Appl. Phys. Lett.*, Vol. 100, p. 101913, 2012.
[228] C. Wan, Y. Wang, N. Wang, W. Norimatsu, M. Kusunoki, and K. Koumoto, Development of novel thermoelectric materials by reduction of lattice thermal conductivity, *Sci. Tech. Adv. Mater.*, Vol. 11, p. 044306, 2010.
[229] C. Wan, Y. Wang, N. Wang, W. Norimatsu, M. Kusunoki, and K. Koumoto, Intercalation: building a natural superlattice for better thermoelectric performance in layered Chalcogenides, *J. Electron. Mater.*, Vol. 40, pp. 1271-1280, 2011.
[230] C. Wan, X. Gu, F. Dang, T. Itoh, Y. Wang, H. Sasaki, M. Kondo, K. Koga, K. Yabuki, and G.J. Snyder, Flexible n-type thermoelectric materials by organic intercalation of layered transition metal dichalcogenide TiS2, *Nat. Mater.*, Vol. 14, pp. 622-627, 2015.
[231] C. Wan, Y. Kodama, M. Kondo, R. Sasai, X. Qian, X. Gu, K. Koga, K. Yabuki, R. Yang, and K. Koumoto, Dielectric mismatch mediates carrier mobility in organic-intercalated layered TiS2, *Nano Lett.*, 2015.

# Tables

Table I. Experimental measurements on the thermal conductivity of graphene

| Ref. | Method | Sample type | Thermal conductivity (W/mK) | Experiment condition and the dependence of thermal conductivity on other factors if reported |
|---|---|---|---|---|
| Balandin [25] | Raman | Exfoliated, pristine, single-layer | 5300±480 ($T$=300K) | 488-nm-wavelength laser, G band, $\chi$= –1.6 $\times 10^{-2}$cm/K, $\delta P_{gra}/\delta P$= 13%, 0.5-to-1-μm-diameter laser spot, 3-μm-wide trench, ambient condition. |
| Ghosh [83] | Raman | Exfoliated, pristine, single-layer | 3080-5150 ($T$=300K) | 488-nm-wavelength laser, G band, $\chi$= –1.6 $\times 10^{-2}$cm/K, $\delta P_{gra}/\delta P$= 11-12%, 1-μm-diameter laser spot, 1-to-5-μm-wide trench, ambient condition. |
| Ghosh [27] | Raman | Exfoliated, pristine, multilayer | 2800->1300 (layer number 2->8) | 488-nm-wavelength laser, G band, 0.5-to-1-μm-diameter laser spot, 1-to-5-μm-wide trench, ambient condition. Thermal conductivity of few-layer graphene decreases with layer number. |
| Faugeras [32] | Raman | Exfoliated, transferred, single-layer | 600 ($T$=660K) | 632.8-nm-wavelength laser, Stokes/anti-Stoke signal, $\delta P_{gra}/\delta P$= 2.3%, 2-μm-diameter laser spot, 44-μm-diameter hole, ambient condition. |
| Cai [30] | Raman | CVD, transferred, single-layer | 2500 ($T$=350K) 1400 ($T$=500K) | 532-nm-wavelength laser, G band, $\chi$= –(4.05±0.2) $\times 10^{-2}$cm/K, $\delta P_{gra}/\delta P$= 3.3±1.1%, 0.34-μm- and 0.56-μm-diameter laser spot, 3.8-μm-diameter hole, ambient condition. |
| Chen [33] | Raman | CVD, transferred, single-layer | 2600-3100 ($T$=350K) | 532-nm-wavelength laser, 2D band, $\chi$= –(7.2±0.2) $\times 10^{-2}$cm/K, $\delta P_{gra}/\delta P$= 3.4±0.7%, 0.34-μm- and 0.54-μm-diameter laser spot, 9.7-μm-diameter hole, in vacuum. |
| Vlassiouk [93] | Raman | CVD, transferred, single-layer | 1200 ($T$=300K) | 514-nm-wavelength laser, G band, $\chi$= –4 $\times 10^{-2}$cm/K, $\delta P_{gra}/\delta P$= 2.3%, 0.4-μm-diameter laser spot, ambient condition. |
| Lee [31] | Raman | Exfoliated, pristine, single-layer | 1800 (T=325K) 710 (T=500K) | 514.5-nm-wavelength laser, 2D band, $\chi$= –(7.2±0.2) $\times 10^{-2}$cm/K, $\delta P_{gra}/\delta P$= 2.3%, 0.58-μm-diameter laser spot, 2.6-μm-, 3.6-μm-, 4.6-μm- and 6.6-μm-diameter holes, ambient condition. |
| Chen [94] | Raman | CVD, transferred, single-layer | 1875 (T=420K, w/o wrinkles) 1480 (T=420K, with wrinkles) | 532-nm-wavelength laser, 2D band, $\chi$= –(7.2±0.2) $\times 10^{-2}$cm/K, $\delta P_{gra}/\delta P$= 2.9±0.2%, 0.34-μm-diameter laser spot, 2.8-μm-diameter holes, ambient condition with the correction of the convection heat lose. |
| Chen [28] | Raman | CVD, transferred, single-layer | 4120±1410 (T=320K, 0.01% $^{13}$C) | 532-nm-wavelength laser, 2D band, $\chi$= –(7.23, 7.05, 6.98, 7.31) $\times 10^{-2}$cm/K for 0.01%, 1.1%, 50% and 99.2% $^{13}$C-labeled graphene, $\delta P_{gra}/\delta P$= 2.9±0.2%, 0.34-μm- |

| | | | 2600±658 (T=330K, 1.1% $^{13}$C) | diameter laser spot, 2.8-μm-diameter holes, ambient condition with the correction of the convection heat loss. Observed U-shape-dependence of the thermal conductivity on the composition. |
|---|---|---|---|---|
| Li [95] | Raman | CVD, transferred, single-layer/bi-layer | 2778 ±569 (T=310 K, single-layer) 1896 ±410 (T=314 K, Bernal stacked bi-layer) 1413 ±390 (T=323 K, twisted bi-layer) | 488-nm-wavelength laser, 2D band, χ= –(5.9, 3.1, 3.4) ×10$^{-2}$cm/K for single-layer, bi-layer AB stacking and twisted graphene, δP$_{gra}$/δP= 3.4%, 6.8% and 6.9% for three samples, 0.316-μm-diameter laser spot, 2.8-μm-diameter holes. The thermal conductivity of twisted bi-layer graphene is smaller than Bernal stacked bi-layer graphene. |
| Wang [41] | Micro-bridge | Exfoliated, transferred, five-layer | 1800 (T=305K) | Electron beam heating; suspended length: 1 μm; sample width: 5 μm. The temperature-dependent thermal conductivity follows $\kappa \propto T^{1.4}$ at low temperature. |
| Pettes [26] | Micro-bridge | Exfoliated, transferred, two-layer | 620±80; 560±70 (T=300K) | The temperature-dependent thermal conductivity follows $\kappa \propto T^{1.5}$ at low temperature. |
| Dorgan [96] | Micro-bridge | CVD, transferred, single-layer; Exfoliated, pristine, single layer | 2500 (T=300K) 310+200/−100 (T=1000K) | Sample dimension: 0.85 μm in width, 1.5 μm in length. The temperature-dependent thermal conductivity follows $\kappa \propto T^{-1.7}$ above room temperature. |
| Wang [97] | Micro-bridge | Exfoliated, transferred, tri-layer | 1400 ±140; 1495 ±150 (T=300 K) | Sample dimension: 5 μm in length, 5.04 μm in width (sample 1); 5 μm in length, 12.8 μm in width (sample 2). The thermal conductivity continuously decreases with increasing coverage of gold nano-particles deposited on the graphene. |
| Xu [88] | Micro-bridge | CVD, transferred, single-layer | 1,689 ±100 (T=300K, L=9 μm) | Sample width 1.5 μm. The length-dependent thermal conductivity follows $\kappa \propto \log(L)$. |

Table II. Theoretical calculations on thermal transport in graphene and graphene nanoribbon.

| Ref | Method | Potential | Room-temperature thermal conductivity (W/mK) | The Dependence of thermal conductivity |
|---|---|---|---|---|
| Lindsay [62] | PBTE-iterative | Tersoff | 3500 ($L$=10 μm) | The important role of ZA modes on thermal conductivity was identified. |
| Singh [90] | PBTE-iterative | Tersoff | 3500 | The thermal conductivity reduces when the layer number of graphene increases from 1 to 4. |
| Lindsay [66] | PBTE-iterative | First-principles | 2897-3155 (naturally occurring, $L$=10 μm) | 14% enhancement is found when the graphene with naturally occurring C isotopes is changed to isotopically pure one; 1% tensile strain leads to a decrease of thermal conductivity. |
| Fugallo [91] | PBTE-iterative | First-principles | 3600 (naturally occurring) | The effects of isotopic disorder, strain and grain boundary are discussed. |
| Gu [54] | PBTE-iterative | First-principles | 3370 (isotopically pure) | The dependence of thermal conductivity on temperature and sample size was reported. |
| Bonini [92] | PBTE-SMRTA-LT | First-principles | 550 | When strain is applied, the thermal conductivity becomes unbounded. |
| Nika [89] | PBTE-SMRTA-Klemens | Valence force field + first-principles | 2000-5000 | The effects of isotopic disorder, sample size and edge roughness are studied. |
| Aksamija [98] | PBTE-SMRTA-Klemens | Valence force field | 5200 (5 μm in width) | The effects of width, edge chirality, edge roughness and temperature on the thermal conductivity of graphene nanoribbon were studied. |
| Aksamija [99] | PBTE-SMRTA-Klemens | Valence force field | 350-1500 | The effects of grain size and grain boundary roughness on the thermal conductivity of polycrystalline graphene were studied. |
| Mei [100] | PBTE-SMRTA-Klemens | Valence force field | 5800 | The effects of width, edge chirality and temperature on the thermal conductivity of graphene nanoribbon were studied. The thermal conductivity of graphene nanoribbon keeps increasing until the width is larger than 100 μm. |
| Evans [101] | EMD | Tersoff | 5000-8000 | The effects of width, edge roughness, and hydrogen termination on the thermal conductivity of graphene nanoribbon were studied. |

| Guo [102] | NEMD | Tersoff | 218 (armchair, 11 nm in length, 2 nm in witdth) 472 (zigzag, 11nm in length, 2 nm in width) | The effects of graphene nanoribbon width, length, edge chirality, and strain effects were studied. |
|---|---|---|---|---|
| Chen [28] | EMD | REBO | 2,859 (pure) 1151 (50% $^{13}$C) | The thermal conductivities of graphene containing different composition of $^{13}$C were calculated. |
| Li [76] | EMD | Terosff | 5500 (unstrained) | The thermal conductivity non-monotonically changes with strain. Maximum value was found in unstrained graphene. |
| Pereira [103] | EMD | Tersoff | 1000 (unstrained) | Thermal conductivity becomes unbounded when a uniaxial strain larger than 2% is applied. |
| Zhang [104] | EMD | REBO | 2900 | The thermal conductivity is reduced to 3 W/mK when graphene has 8.25% vacancy defects. |
| Haskins [105] | EMD | Tersoff | 2300 | The thermal conductivity is reduced to 20% when graphene nanoribbon has 0.1% single vacancy concentration or 0.23% double vacancy or Stone-Wales concentration. |
| Mingo [106] | AGF | First-principles | - | The phonon transport of graphene with isotope cluster was studied in ballistic regime. |
| Mortzavi [107] | EMD | REBO | 280 | The thermal conductivity gradually decreases to 3% of pristine graphene when the grain size is 1nm. |
| Liu [108] | EMD | Tersoff | 2430 | The thermal conductivity decreases exponentially with increasing GB energy. |

Table III. Experimental and theoretical studies on thermal transport in h-BN.

| Experiment: | | | | |
|---|---|---|---|---|
| Ref. | Method | Sample type | Room-temperature thermal conductivity (W/mK) | Experiment condition and the dependence of thermal conductivity if reported |
| Jo [39] | Micro-bridge | exfoliated, transferred | 250 (5-layer) 360 (11-layer) | Dimension of the sample: 7.5 μm in length; vacuum. |
| Zhou [114] | Raman | CVD, transferred | 227-280 (9-layer) | 2G band, $\chi = -(3.78 \pm 0.16) \times 10^{-2}$ cm/K, $\delta P_{h-BN}/\delta P = 10\%$, 1-μm-diameter laser spot, 7-μm-wide trench, ambient condition. |
| Simulation: | | | | |
| Ref | Method | Potential | Room-temperature thermal conductivity (W/mK) | The dependence of thermal conductivity |
| Sevik [110] | EMD | Tersoff | 400 | Zigzag nanoribbons are of larger thermal conductivity than the armchair ones at the same width; The thermal conductivities of both types of nanoribbons converged to bulk value. |
| Lindsay [111] | PBTE-iterative | Tersoff | ~600 ($L$=2 μm) | The thermal conductivity increases from 500W/mK to 800 W/mK when the sample size is increased from 1μm to 10 μm; strong thermal conductivity enhancement is found when converting the h-BN with naturally-occurring B and N isotopes to isotopically pure h-BN. |
| Lindsay [112] | PBTE-iterative | Tersoff | ~600 ($L$=2 μm) | Thermal conductivity of few-layer h-BN decreases with the increase of the layer number. |
| Ouyang [115] | AGF | Valence force-field | 1700-2000 | The thermal conductance depends linearly on the width; the thermal conductance follows $T^{1.5}$. |
| Yang [116] | AGF | Valence force-field | - | The thermal conductance of nanoribbons decreases linearly with the size increase of triangle vacancies |

Table IV. Theoretical calculations on the thermal transport in silicene.

| Ref | Method | Potential | Room-temperature thermal conductivity (W/mK) | Thermal conductivity dependence |
|---|---|---|---|---|
| Gu [54] | PBTE-iterative | First-principles | 26 (L=10 μm) | The thermal conductivity logarithmically increases with sample size. |
| Xie [123] | PBTE-SMRTA-LT) | First-principles | 9.4 | The mode-dependent thermal conductivity was reported |
| Hu [127] | NEMD | Tersoff | 40 | The thermal conductivity increases with the uniaxial tensile strain. |
| Pei [126] | NEMD | MEAM | 30.5 | The thermal conductivity exhibits a U-shape dependence with the composition of the doped $^{32}$Si; The thermal conductivity shows a maximum value when a 4% tensile strain is applied. |
| Li [128] | EMD | Tersoff | 15 | The dependence of thermal conductivity on vacancy defects with different size, shape and concentration was reported. |
| Wang [129] | NEMD | Tersoff | 55 | A non-U-shape dependence of thermal conductivity on the composition of carbon was found when silicon atoms are substituted by carbon atoms. |
| Pan [130] | AGF | Tersoff | - | The thermal conductance increase with the increase of ribbon width. |
| Zhang [49] | EMD/NEMD | Stillinger-Weber | 5.5-11.77 | Size effects on thermal conductivity were studied. |

Table V. Experimental and theoretical studies on thermal transport in $MoS_2$.

| \multicolumn{5}{c}{Experiment:} | | | | |
|---|---|---|---|---|
| Ref. | Method | Sample type | Room-temperature thermal conductivity (W/mK) | Experimental conditions and the dependence of thermal conductivity if reported |
| Sahoo [29] | Raman | CVD, transferred | 52 (11-layer) | $A_{1g}$ band, $\chi = -1.23 \times 10^{-2}$ cm/K, $\delta P_{h-BN}/\delta P = 10\%$, 1-1.5 μm laser spot |
| Yan [137] | Raman | exfoliated, transferred | 34.5 (monolayer) | $A_{1g}$ mode, $\chi = 0.011$ cm$^{-1}$/K, $\alpha = 9 \pm 1\%$, 170 nm diameter laser spot, 1.2-μm-diameter hole, ambient condition |
| Jo [138] | Micro-bridge | exfoliated, transferred | 44–50 (4-layer) 48-52 (7-layer) | Suspended length: 3 and 8 μm, sample width: 5.2 and 2.2 μm. |
| Taube [139] | Raman | exfoliated, pristine | 62 (supported monolayer) | $A_{1g}$ mode, ambient condition |
| \multicolumn{5}{c}{Simulation:} | | | | |
| Ref | Method | Potential | Room-temperature thermal conductivity (W/mK) | The dependence of thermal conductivity |
| Li [67] | PBTE-SMRTA-LT | First-principles | 91 (L=1 μm) | The thermal conductivity of isotopically pure $MoS_2$ is 10% larger than $MoS_2$ with naturally occurring Mo and S isotopes. |
| Gu [55] | PBTE-iterative | First-principles | 103 (L=1 μm) | Strong length dependence is found in $MoS_2$. |
| Cai [59] | PBTE-SMRTA-Klemens | First-principles | 23.3 | - |
| Jiang [50] | NEMD | Stillinger-Weber | 5.5 | Thermal conductivity is reduced to 60% when a 8% tensile strain is applied. |
| Liu [142] | NEMD | Bond, angle, Coulomb | 1.35 | Thermal conductivity of $MoS_2$ nanoribbons is insensitive to width and edge-type. |
| Muratore [143] | NEMD | Harmonic bonds and angle + Lennard-Jones + Coulomb interaction | 44 | Thermal conductivity increases from 1.3 to 44 when the length increases from 5 to 70 nm. |

Table VI. Theoretical calculations on thermal transport in graphene derivatives.

| Ref | Material | Method | Potential | Room-temperature thermal conductivity (W/mK) | Thermal conductivity reduction compared with pristine graphene |
|---|---|---|---|---|---|
| Fugallo [91] | Graphene | PBTE-iterative | First-principles | 1860 (L=100 μm) | 50% |
| Kim [215] | Graphane | EMD | REBO | 177 | 46% |
| Pei [214] | Graphane | NEMD | AIREBO | ~30 | 33% |
| Liu [218] | Graphane | NEMD | AIREBO | 33.2 | 43% |
| Fugallo [91] | Flurographene | PBTE | First-principles | 260 (L=100 μm) | 7% |
| Huang [217] | Flurographene | NEMD | Universal force-field | 40 (L~20 nm) | 35% |
| Mu [154] | Graphene oxide | NEMD | REBO | 8.8 (20% coverage) | 5% |